\documentclass[usenatbib]{mnras}
\usepackage{epsfig}
\usepackage{amsmath,amssymb}
\usepackage{color}
\usepackage{booktabs}
\usepackage{pdflscape}

\newcommand{\Msolar}{\mbox{\,$\rm M_{\odot}$}}        
\newcommand{\Rsolar}{\mbox{\,$\rm R_{\odot}$}}        
\newcommand{\kms}{\mbox{\,$\rm km\,s^{-1}$}}        
\newcommand{\perday}{\mbox{\,$\rm d^{-1}$}}        
\newcommand{\teff}{\mbox{\,$T_{\rm eff}$}}      
\newcommand{\logg}{\mbox{\,$\log g$}}        
\newcommand{\lgcs}{\mbox{\,$\log g / {\rm cm\,s^{-2}}$}}        
\newcommand{\vt}{\mbox{$v_{\rm t}$}}             
\newcommand{\vrot}{\mbox{$v_{\rm rot} \sin i$}}             
\newcommand{\nh}{\mbox{\,$n_{\rm H}$}}        
\newcommand{\nhe}{\mbox{\,$n_{\rm He}$}}        
\newcommand{\ew}{\mbox{$W_{\lambda}$}}               
\newcommand{\lineA}[3]{\mbox{#1{\sc #2}\,$\lambda #3$\AA}}        

\newcommand{\ledz}{UVO\,0825+15}   
\newcommand{\crimson}{LS\,IV$-14^{\circ}116$}  

\newcommand{\rooster}{HE\,2359--2844}  
\newcommand{\kraft}{HE\,1256--2738}  
\newcommand{\quartz}{UVO\,0512-08}  
\newcommand{\ethel}{PG\,0909+276} 

\title[UVO 0825+15: a variable lead-rich hot subdwarf]{Discovery of a variable lead-rich hot subdwarf:  UVO 0825+15}

\author[C.S. Jeffery et al.]
       {C. S. Jeffery,$^{1,2}$\thanks{E-mail: csj@arm.ac.uk}
       A. S. Baran,$^3$
       N. T. Behara,$^{1}$ 
      A. Kvammen,$^{4}$
      P. Martin,$^{1,2}$ 
      \newauthor
       Naslim N.,$^{5,6}$
      R. H. \O{}stensen,$^7$ 
      H. P. Preece,$^{1,8}$ 
     M. D. Reed,$^{7}$ 
     J. H. Telting,$^{4}$ 
      \newauthor
     and
     V. M. Woolf\,$^{9}$ \\
$^{1}$Armagh Observatory and Planetarium, College Hill, Armagh BT61 9DG\\
$^{2}$School of Physics, Trinity College Dublin, College Green, Dublin 2, Ireland\\
$^{3}$Suhora Observatory and Krak\'ow Pedagogical University, ul. Podchor\c{a}\.zych 2,30-084,Krak\'ow, Poland\\
$^{4}$Nordic Optical Telescope, Rambla Jos\'e Ana Fern\'andez P\'erez 7, 38711 Bre\~na Baja, Spain\\
$^{5}$Academia Sinica Institute of Astronomy and Astrophysics, Taipei 10617, Taiwan, Republic of China\\
$^{6}$National Astronomical Observatory of Japan, 2-21-1 Osawa, Mitaka, Tokyo 181-8588, Japan\\
$^{7}$Department of Physics, Astronomy and  Materials Science, Missouri State University, 901 S. National, Springfield, MO 65897, USA\\
$^{8}$Institute of Astronomy, University of Cambridge, Madingley Rd., Cambridge, CB3 0HA, UK\\
$^{9}$Department of Physics, University of Nebraska at Omaha, 6001 Dodge Street, Omaha, NE 68182, USA\\
}

\date{Accepted .....
      Received ..... ;
      in original form .....}

\pagerange{\pageref{firstpage}--\pageref{lastpage}}
\pubyear{2016}

\begin{document}

\label{firstpage}

\maketitle

\begin{abstract}
\ledz\ is a hot bright helium-rich subdwarf which 
lies in {\it K2} Field 5 and in a sample of intermediate
helium-rich subdwarfs observed with {\it Subaru/HDS}. The {\it K2}
light curve shows low-amplitude variations, whilst 
the {\it Subaru} spectrum shows Pb{\sc iv} absorption lines, indicative of  a very 
high lead overabundance.  \ledz\ also has a high proper motion with kinematics typical 
for a thick disk star.  Analyses of ultraviolet and intermediate
dispersion optical spectra rule out a short-period  binary companion, and 
provide fundamental atmospheric parameters of $\teff=38\,900\pm270$\,K, $\lgcs=5.97\pm0.11$, 
 $\log n_{\rm He}/n_{\rm H}=-0.57\pm0.01$, $E_{B-V}\approx0.03$, 
and angular radius $\theta = 1.062\pm0.006\times10^{-11}$ radians (formal errors). 
The high-resolution spectrum shows that carbon is $>2$ dex subsolar,  iron is approximately solar
and  all other elements heavier than argon are at least 2 -- 4 dex overabundant, including 
germanium, yttrium and lead. Approximately 150 lines in the blue-optical spectrum 
remain unidentified.  The chemical structure of the photosphere is presumed to be determined by 
radiatively-dominated diffusion. The {\it K2} light curve shows a dominant period around 10.8\,h,  with a
variable amplitude,  its first harmonic, and another period at 13.3\,h. The preferred explanation 
is multi-periodic non-radial oscillation due to g-modes with very high radial order, although this
presents difficulties for pulsation theory. Alternative explanations fail for lack 
of radial-velocity evidence.  \ledz\ represents the fourth member of a group of hot subdwarfs having helium-enriched
photospheres and  3--4 dex overabundances of  trans-iron elements, and is the first
lead-rich subdwarf to show evidence of pulsations. 
\end{abstract}



\section{Introduction}

The theoretical helium ``main-sequence'' is defined by 
stars consisting  almost entirely of helium, and converting helium to carbon at their centres via the nuclear
triple-alpha process, which form a continuum that depends primarily upon the mass of the star. 
There exists a diverse population of low-mass hot stars which have surface properties
placing them close to this sequence. They are commonly referred to as ``hot subdwarfs'', since their 
surfaces are between 4 and 8 times hotter than that of the Sun,  and they lie below the more familiar 
hydrogen main-sequence,  where stars are referred to as ``dwarfs'' to distinguish them 
from the larger  giants and supergiants.  In contrast to the surfaces of hydrogen main-sequence stars,
which persistently show a helium:hydrogen ratio of around 1:9 (by  number),   
hot subdwarfs show considerable diversity with {\it surface} helium:hydrogen ratio  ranging 
from the extremely helium-poor subdwarf B  (sdB) stars (1:10000), 
extremely helium-rich subdwarf O (He-sdO) stars (99:1), and a 
small number of intermediate composition (1:9 -- 9:1). This diversity reflects a 
variety of evolutionary origin evidenced by various duplicities, with sdB stars having 
white dwarf, M dwarf, F-G dwarf, or planetary companions, or none. Progress in 
understanding these origins has been assisted by the discovery of pulsations in 
both sdB and sdO stars, although these are not universal. A recent review examines
many of the properties of hot subdwarfs, and the problems surrounding them \citep{heber16}. 

The helium-rich subdwarfs themselves represent a heterogeneous group having a  wide range in 
surface temperature ($25\,000 - 45\,000$\,K) \citep{naslim10}, including at least one 
spectroscopic binary \citep{naslim12}, and a small group with 
extraordinary (3 -- 4 dex) overabundances of germanium, strontium, yttrium, zirconium,   
and lead \citep{naslim11,naslim13}. It has been assumed
that radiative levitation selectively lifts certain species into the line-forming region in the atmospheres of
these stars; a proof that this explanation is correct is still wanting.
The prototype of these ``heavy-metal stars'', the zirconium star   \crimson\ 
pulsates with multiple periods of around 1800\,s \citep{ahmad05, green11}.  No instability
mechanism has yet been  identified by which to drive these pulsations. 

This paper is concerned with two questions; 1) evidence for pulsations in hot subdwarfs which fall within 
the fields observed by the spacecraft {\it K2} and 2) the origin of the class of intermediate helium-rich 
subdwarfs. By coincidence, the same star was chosen to address each of these questions with
 observations carried out during 2015 in  {\it K2} (Campaign 5) and  in a high-resolution 
spectroscopic survey with the Subaru telescope on Mauna Kea.  

The star, known variously as  TD1 31206 = TD1 32707 \citep{thompson78} =
UVO 0825+15 \citep{berger80} = [CW83] 0825+15 \citep{carnochan83} = TYC 808-490-1 =  
GALEX J082832.8+145205 =  2MASS J08283287+1452024
= EPIC 211623711, 
was classified ``sdO" by \citet{berger80} and \citet{kilkenny88}. 
The equatorial coordinates for epoch J2000 and equinox 2000  
are $\alpha = 08h 28m 32.876s$, $\delta = +14^{\circ} 52' 02.51''$. The {\it SIMBAD} 
database \citep{wenger00} does not associate \ledz\ with that identified by other synonyms 
due to low positional precision. The star exhibits a substantial proper motion, exceeding 24 mas\,y$^{-1}$ 
\citep{hog00}. 

Observations with GALEX and the NTT were used by
\citet{vennes11} to derive an effective temperature $T_{\rm eff}=36\,650\pm650$\,K, a surface gravity $\lgcs =5.65\pm0.14$, 
and a surface helium-to-hydrogen ratio $\log n_{\rm He}/n_{\rm H} = -0.50\pm0.08$.  \citet{nemeth12} used 
low-dispersion optical spectra to obtain 
$T_{\rm eff}=37\,060\pm600$\,K, $\lgcs =5.92\pm0.10$, and  $\log n_{\rm He}/n_{\rm H} = -0.62\pm0.08$. 

In the following, we describe the observations (\S 2),  
the derivation of photospheric properties (\S 3) and discovery of strong lead
absorption lines (\S 4), the discovery of light variability, 
analysis of the light curve and its interpretation (\S 5),
and analysis of the kinematics (\S 6). 
We compare these properties with those of other helium-rich subdwarfs (\S 7). 


\section{Observations}
\label{s:obs}

\subsection{Spectrophotometry: {\it IUE} and $BVRJHK$}
\label{s:uv_obs}
As an ultraviolet bright source identified in the {\it TD1A S2/68} ultraviolet
sky survey \citep{carnochan83}, \ledz\ was an early target for the 
{\it International Ultraviolet Explorer (IUE)}. Images LWR09914 and SWP11306 were 
obtained using both small and large apertures on 1981, February 14. We have 
used the large-aperture images for photometric reliability, supplemented by the 
small-aperture data in the range 1200--1400\AA\ where many large-aperture
pixels are saturated.

In addition there exists photometry in broad-band filters. From the
{\it SIMBAD} database, 
$B =11.57 \pm 0.10$, 
$V =11.82 \pm0.21$,   
$R =12.01 \pm0.04$, 
$J =12.425 \pm0.021$, 
$H =12.586 \pm0.026$, and
$K= 12.652 \pm0.029$ mag
\citep{hog00,zacharias09,cutri03}.

\begin{table}
\caption{Relative radial velocities of \ledz\ from {\it NOT/ALFOSC} spectroscopy.}
\label{t:not_rv}
\begin{tabular}{lrr}
BJD               &  $\delta v$ & $\pm$  \\               
2457436.3442983 &   9.94 & 3.62     \\ 
2457443.5281802  & $-$12.27 & 5.64     \\
2457444.5318080 &   5.36 & 3.94      \\
2457447.5096600  & $-$9.54 & 5.05     \\
2457447.5833238 & $-$12.28 & 4.41   \\
2457449.4176574 &  $-$5.46 & 4.77  \\
\end{tabular}
\end{table}

\subsection{{\it Nordic Optical Telescope/ALFOSC} spectroscopy}
\label{s:not_obs}
We obtained 6  low resolution spectra between 2016 February 17 and March 1 
using the 2.56-m Nordic Optical Telescope with ALFOSC, grism \#18 and 
a 0.5" slit, with exposure times of 150--200\,s. 
The spectra have  $R \approx2000$ (resolution element 2.2\AA) 
and S/N  ranging from 55 to 130.

The data were homogeneously reduced and analysed. Standard reduction steps 
within {\sc iraf} include bias subtraction, removal of pixel-to-pixel 
sensitivity variations, optimal spectral extraction, and wavelength 
calibration based on arc-lamp spectra. The target spectra and the 
mid-exposure times were shifted to the barycentric frame of the solar 
system.

Radial velocities were derived with the {\sc fxcor} package in {\sc iraf}. 
We used the  H\,$\beta$, H\,$\gamma$, H\,$\delta$, H\,$\zeta$ and H\,$\eta$ lines 
to determine the radial velocities (RVs),  and used the normalised average spectrum as a template. 
The final RVs were  adjusted for the position of the target in the slit, judged from slit 
images taken just before and just after the spectral exposure. 

The relative velocities are shown in Table \ref{t:not_rv}, where the root mean square 
deviation around the mean of 8.7\kms\ is less than twice the mean standard 
error (4.6\kms, individual errors from {\sc fxcor}). With six measurements 
obtained over 2 weeks, this rules out any short-period ($< 7$ d) binary 
companion unless the inclination is very small.

\subsection{{\it Subaru/HDS} spectroscopy}
\label{s:hds_obs}
\ledz\ was observed with the High Dispersion Spectrograph (HDS)  \citep{noguchi02} on the {\it Subaru} 
telescope, operating in service mode, on 2015 June 3. Two observations were made consecutively, 
each with exposure time 240\,s. 
A slit width of 0.4\,mm was used, corresponding to a projected  resolution $R=45\,000$. 
 The data were reduced using standard {\sc iraf} procedures, together with the reduction manual 
and scripts written by {\it Subaru} staff for reducing HDS spectra \citep{aoiki08,tujitsu13}.
Wavelengths 
were corrected for earth motion, the two observations were combined, and the orders 
were merged to provide a single rectified spectrum. Instrumental artefacts resembling broad aborption
lines with emission wings (or vice versa) occur in a number of places. These were clipped 
or masked from the spectrum during analysis. 

The radial velocity for the combined order-merged spectrum was
measured using cross-correlation against two theoretical spectra of differing 
\teff\ and metallicity and using two spectral ranges, one in each part of the
spectrogram. These four measurements gave a mean radial 
velocity of $+56.4\pm0.5$\kms.

\begin{figure*}
\epsfig{file=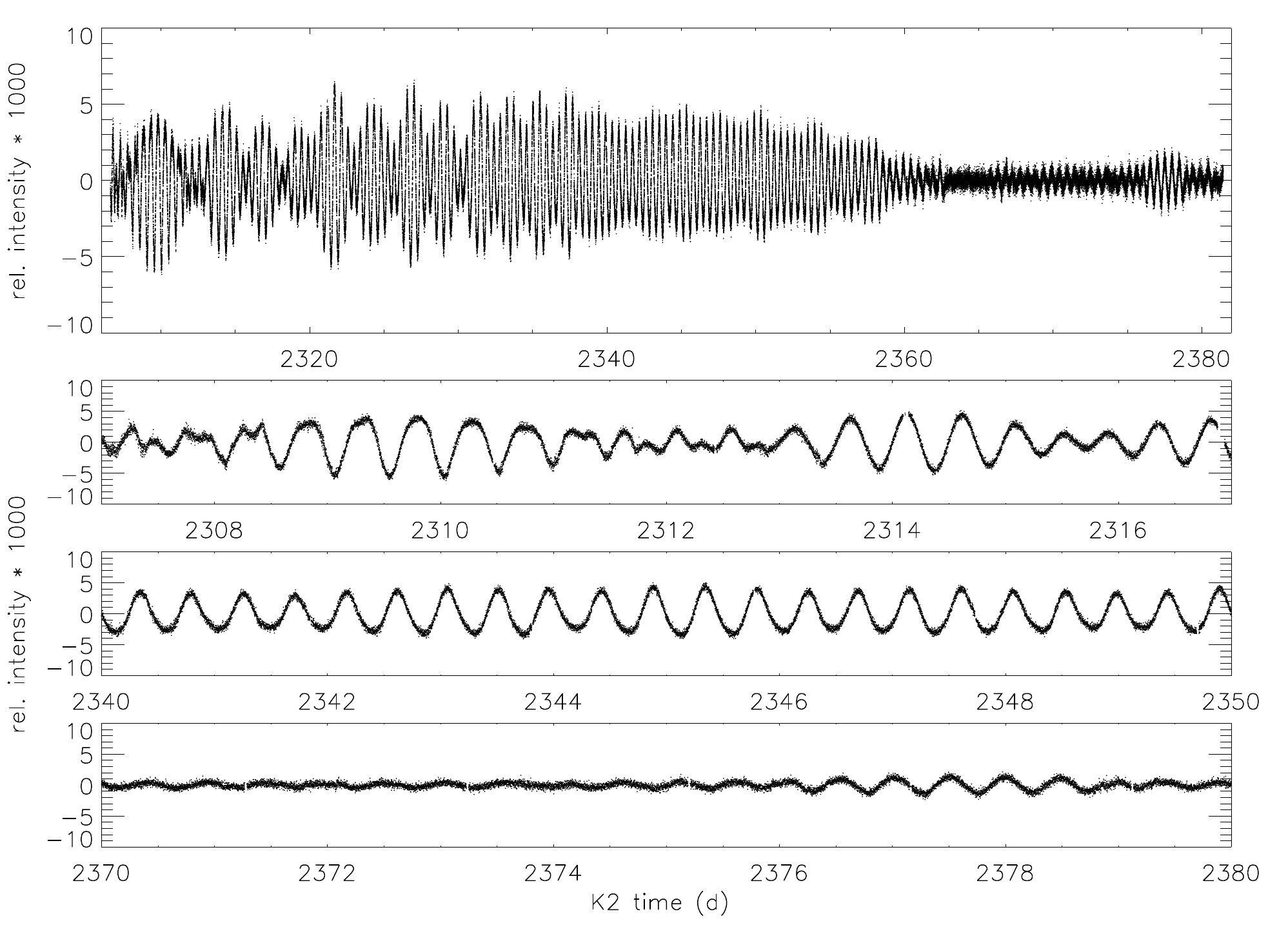,width=175mm,angle=0}
\caption{{\it K2} short cadence light curve of \ledz\ showing the entire dataset from Campaign 5 (top) and
three 10-day segments (below) displaying different characteristics.  } 
\label{f:k2}
\end{figure*}

\subsection{{\it K2} photometry}
\label{s:k2_obs}
Short cadence {\it K2} observations, providing one image every 58.85\,seconds, were obtained between 2015 April 27 and 2015 July 10. 
The \emph{K2} mission only uses two gyros for stability and so the spacecraft
rolls from solar pressure, requiring occasional thruster firings for pointing
corrections. Consequently, objects drift across an array of pixels and unequal 
pixel sensitivity produces fluxes which are pixel-correlated.
To correct for this  we have developed a processing pipeline using
a combination of {\sc iraf} photometry packages and custom decorrelation routines.
We downloaded pixel files from the Mikulski Archive for Space Telescopes and
used {\sc daofind} to determine the motion of the stellar profile. We then
extracted fluxes using aperture photometry and iteratively decorrelated 
the fluxes in both directions of the pixel array. The stellar signal has three distinct
regions and because the first two are so strong, we found it necessary to prewhiten the
stellar signal prior to determining pixel-flux correlations for those regions.
Pixel-correlated fluxes were then subtracted from the light curve, effectively
removing motion-induced variations. The light curve was then sigma clipped to
remove outliers and spline fitted to remove trends on the order of six days and
longer. The resulting light curve is shown in  Fig.~\ref{f:k2}, where 
 {\it K2} times are defined as barycentric Julian Date (BJD) -- 2454833.0.


\section{Spectroscopic Analysis}

\subsection{Model Atmospheres}

The analysis of stellar spectra depends heavily on the computation of physically
realistic models for the temperature and density structure of  the stellar atmosphere 
and for the detailed spectrum of radiation emergent therefrom. For the majority of this study
we have computed bespoke models for both using Armagh codes {\sc sterne}
\citep{behara06} and {\sc spectrum}\footnote{\tt www.arm.ac.uk/$\sim$csj/software\_store/guide/spectrum.html}, 
respectively. 

{\sc sterne} assumes a plane-parallel atmosphere in radiative, hydrostatic and local-thermodynamic equilibrium (LTE). 
Monochromatic continuous opacities are obtained from the Opacity and Iron Projects \citep[][et seq.]{OP95,hummer93}
for up to six ions of all elements hydrogen through silicon, sulphur, argon, calcium and iron. 
Line blanketing is treated by Opacity Sampling in a list of 559\,000 absorption lines (including all 
species from hydrogen to bismuth, thorium and uranium).
Use of this code allows the speed and flexibility necessary to produce multiple model grids of 
arbitrary composition. Since the input composition needs to match the measured composition of the star, 
which is not known {\it a priori}, some iterations are necessary to obtain a final solution. We note that 
{\sc sterne}  differs from many ``LTE'' codes by allowing for electron scattering in the radiative 
transfer equation. The full LTE approximation, which equates the source function to the mean intensity 
and to the  Planck function (i.e. $S_{\nu}=J_{\nu}=B_{\nu}$), breaks down  at high temperature and 
low surface gravities.  
{\sc sterne}  also provides  emergent fluxes ($F_{\lambda}$) at low resolution from 220 \AA\ to 20 $\mu$m.

{\sc spectrum} is a  formal solution code providing a high-resolution emergent spectrum for
a given input model atmosphere. For consistency, it must use the same continuous opacity and
equation of state as used for the input model. Line profiles for metal lines are computed using
Voigt profiles, including thermal, micro-turbulent, radiative, electron and van der Waal's broadening,
where appropriate line data exist. Stark broadened profiles for H, He$^0$ and He$^+$ lines are computed 
using tables from \citet{vidal73}, \citet{beauchamp97} and \citet{schoning89}.

\begin{figure*}
\epsfig{file=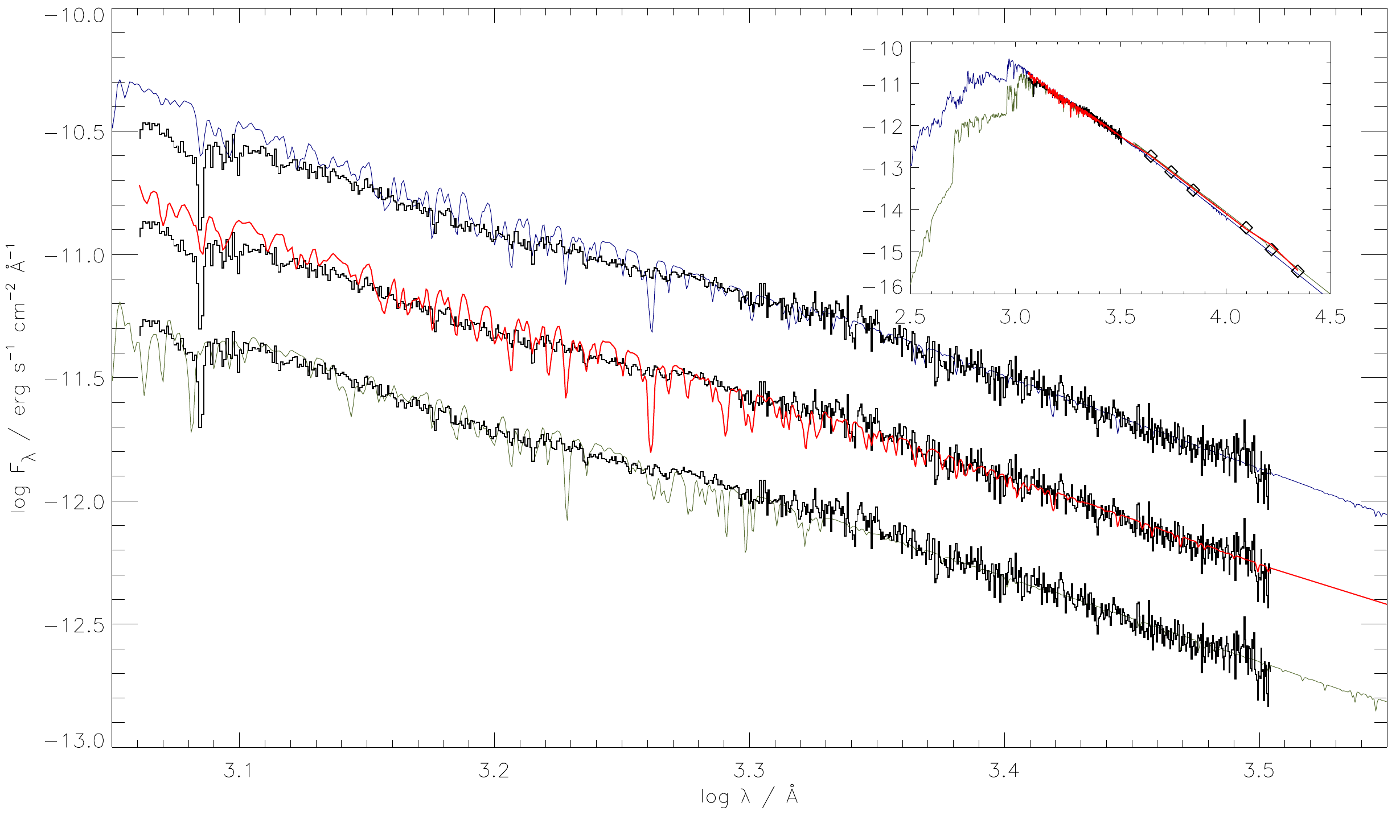,width=170mm,angle=0}
\caption{The merged {\it IUE} spectrum of \ledz\ (black histogram) 
compared with a theoretical spectrum (red),  with properties matched to those 
obtained from optical  intermediate and high-resolution spectroscopy, 
with  $T_{\rm eff}=39\,000$\,K,  $\lgcs=6 $, from grid {\bf h80he20\_uvo0825}.
$E_{\rm B-V}=0.03$ was found by inspection and $\theta=1.062\pm0.006\times10^{-11}$ 
radians is obtained by $\chi^2-$minimization. 
The fit is resampled to the wavelength grid of the observations and 
convolved with an instrumental profile having full-width half-maximum of 7\AA.
Model atmospheres with $T_{\rm eff}=30\,000$ and $45\,000$\,K 
are shown in green and blue and displaced down and up by 0.4 dex, respectively, also
convolved with the instrumental profile, and normalized to the $39\,000$\,K
solution at $V$. The inset panel shows the same data extended
to the far ultraviolet and to the infrared, with $BVRJHK$ magnitudes 
shown as diamonds. } 
\label{f:iue}
\end{figure*}

\subsection{Effective temperature from spectrophotometry}
\label{s:fluxes}

The most reliable way to measure the effective temperature of a hot star is to use the
total-flux method. Ideally the observed bolometric flux ($f = L/4\pi d^2$) is obtained by 
integrating over all wavelengths,  the angular diameter ($\theta = r/d$) is obtained by 
normalising the fluxes at some reference  wavelength to the fluxes of a model stellar 
atmosphere with similar effective temperature,  and the effective temperature deduced
by eliminating $d$ to obtain 
\[\sigma T_{\rm eff}^4 = (f  4\pi d^2 /4 \pi  r^2) = (f / \theta^2),\] 
where $\sigma$ is the Stefan-Boltzmann constant.  Where 
significant parts of the overall flux distribution lie outside the observed wavelength range,
these can be supplied by reference to a suitable model stellar atmosphere. The observed fluxes 
must also be corrected for any interstellar extinction  (de-reddened). Effectively, the
flux integral and dereddening can be substituted by finding the model atmosphere and
extinction curve which best fits the observed fluxes, and deducing $T_{\rm eff}$ and
the colour excess $E_{B-V}$ therefrom. 

There are hazards in this approach: i) the flux {\it distribution} 
(not the {\it total} flux) also depends on the surface gravity and composition and ii)  for hot
stars, the effects of extinction and temperature on the gradient of the ultraviolet
continuum are partially degenerate, except around the broad interstellar 2175\AA\ absorption feature.
Fortunately, in the case of \ledz, there is no evidence for any feature at  2175\AA, 
so we initially adopted $E_{B-V} \approx 0$.  So it only remains to establish appropriate values for
surface gravity and composition. This is an iterative process which requires information
from spectroscopy. 

As a first approximation we computed a grid of line-blanketed model atmospheres with
H:He ratios in the range 100:00, 90:10, 70:30, 50:50, 30:70, 10:90, 1:99, and 
00:100 (labelled {\bf h}{\it xx}{\bf he}{\it yy}, where {\it xx} and {\it yy} correspond 
to the hydrogen:helium number fractions {\it per cent} just described). 
The chemical composition due to all other elements was taken to be solar by number 
(labelled {\bf p00}).  
The micro-turbulent velocity was fixed at  $v_t=5\kms$, typical for B-type 
stars. A test with a small model grid having $v_t=2\kms$ showed no discernible
difference in the theoretical flux distribution, or in the photospheric solution discussed
below.  

The intermediate-dispersion spectroscopy (\S \ref{s:atmos}) indicates $\lgcs \approx 6$ and
$\nh:\nhe \approx 80:20$, so we initially used the grid of atmospheres identified as 
{\bf h70he30p00/t...g600p00}, where {\bf t...} represents $T_{\rm eff}$ in the grid 
(28, \ldots (2) \ldots, 42, 45)\,kK.  

Applying our $\chi^2$-minimization method \citep[\sc ffit: \rm][]{jeffery01a} to 
the {\it IUE}+$BVRJHK$ spectrophotometry
described in \S\,\ref{s:uv_obs}, and after iterating with \S\,\ref{s:atmos}, we were unable
to find a unique solution for $T_{\rm eff}$, $E_{B-V}$ and $\theta$.  
Fig. \ref{f:iue} demonstrates that for this star, the overall  flux distribution is 
insensitive to $T_{\rm eff}$  within the optical and {\it IUE} range, with only the detailed line 
spectrum at high-resolution being affected. It will be necessary to make ultraviolet
observations at shorter wavelength or at higher resolution in order to break this degeneracy. 
We also note that the models show insufficient absorption around $\lambda1215$\AA. This is 
due to the omission of Stark broadening from the \ion{He}{ii} series in the calculation of the model
atmosphere; the hydrogen Ly$\alpha$ line is very weak at the \teff of the solution. This omission 
and the use of only 559034 strong lines in the input line list introduces a small systematic error 
to the predicted flux distribution. However this error is small compared with the effects of both line 
and continuous far-ultraviolet opacity due to other ions, such as calcium, when significantly 
overabundant. 

Consequently, it was necessary to adopt $T_{\rm eff}=39\,000$\,K 
on the basis of medium-resolution optical  spectroscopy (\S \ref{s:atmos}).
The procedure was repeated after obtaining a first estimate for the surface composition (cf. \S~\ref{s:abunds})
and computing a model grid based on this mixture (labelled {\bf h80he20\_uvo0825}).
Some experimentation showed that $E_{B-V}$ could not be exactly zero; a value of 
0.03 was obtained by inspection. $\theta=1.062\pm0.006\times10^{-11}$\,rad 
is given by fitting the model to the observations  (Fig.~\ref{f:iue}). 
The latter is subject to a systematic error of $\pm10\%$ arising from 
the possibility of unknown extreme- and far-ultraviolet opacity arising from 
abundant species  ($Z>26$) not included in the model (i.e. not included in 
either the Opacity or Iron Projects). 

\begin{figure*}
\epsfig{file=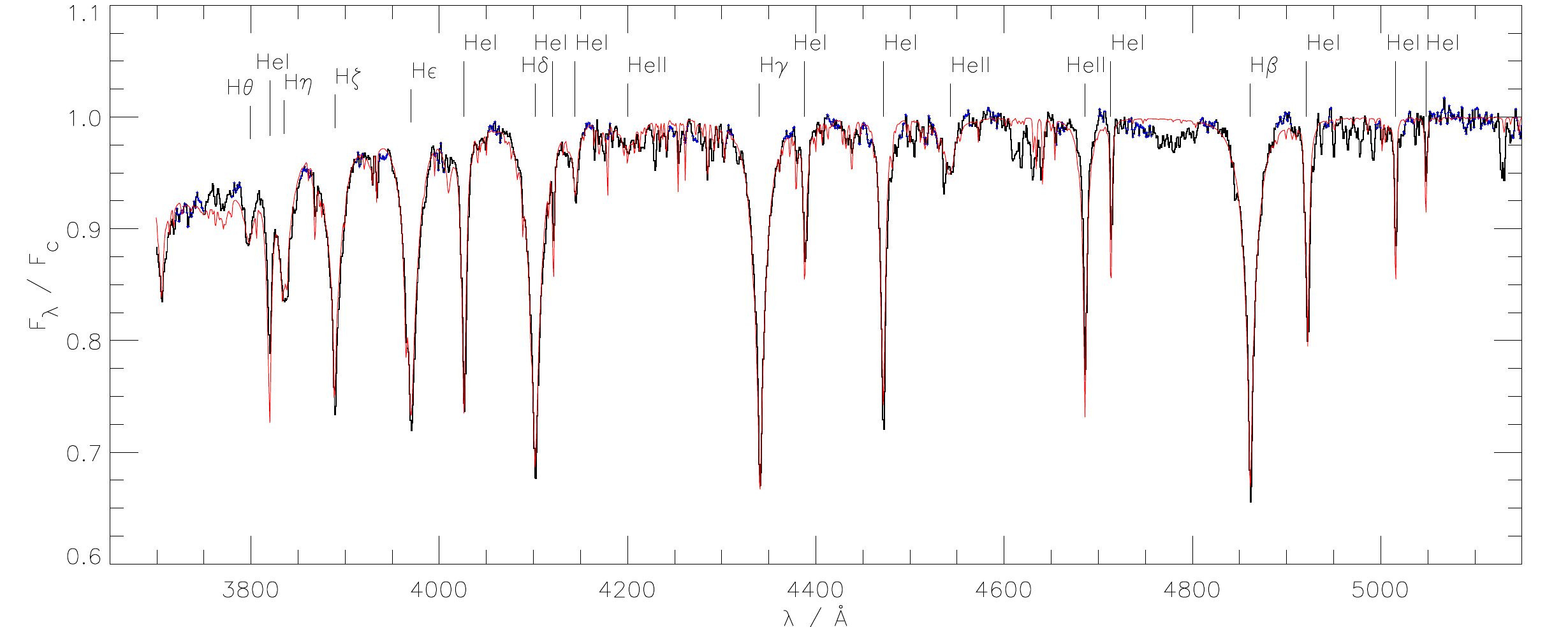,width=170mm,angle=0}
\caption{The mean {\it NOT/ALFOSC} spectrum of \ledz\ (black histogram) 
compared with the best-fit spectrum (red), obtained from a grid of line-blanketed 
model atmospheres with composition tailored to the observations and having the solution
$T_{\rm eff}=38\,900\pm270$\,K,  $\lgcs=5.97\pm0.11$, $\nhe = 0.21\pm0.05$
 ($\log n_{\rm He}/n_{\rm H} = -0.6\pm0.01$). Pseudo-continuum 
points used for re-normalisation are marked 
in blue. } 
\label{f:alfosc}
\end{figure*}

\subsection{Surface gravity and hydrogen:helium ratio from {\it NOT/ALFOSC} spectroscopy}
\label{s:atmos}

A second option for securing \teff\ and, in addition, \logg, and \nhe/\nh, is to use the profiles of 
hydrogen, and neutral and ionized helium lines in the blue-optical spectrum. The widths of these lines 
in hot stars are sensitive to pressure and hence to \logg; \nhe/\nh can be measured 
from the relative strength of  hydrogen and helium lines;  \teff\ can be measured from 
the neutral-to-ionized helium ratio as well as from the Balmer decrement.

Our approach was to seek an optimum fit to the mean {\it NOT/ALFOSC} spectrum of \ledz\ 
using $\chi^2$-minimisation in a grid of models  as follows.
The entire wavelength region between 3680 and 5150\AA\ was used. 
The observed spectrum was re-normalized prior to each $\chi^2$-minimisation 
using a pseudo continuum obtained as follows.
Regions of spectrum deemed to be free from 
both observed and predicted absorption lines were identified as pseudo-continuum regions.
A predicted spectrum, resampled onto the wavelength grid of the observed spectrum,
 was divided throughout by the latter. The errors associated with each wavelength
in the pseudo-continuum regions of the observed spectrum were associated 
with the ratioed spectrum.  Elsewhere, the errors were set to $10^{10}$.
The ratioed spectrum was smoothed, using the inverse of the associated errors 
to weight individual data points and by convolution with a Gaussian having full-width 
half-maximum of 200\AA.  
The observed spectrum was  multiplied by this smoothed 
pseudo-continuum to obtain a re-normalized spectrum. The object was to 
ensure a satisfactory fit even in regions of spectrum where overlap between
broad lines means there is no true continuum, and also to ensure that the renormalisation
does not affect the local profile of any individual absorption line.  
A $\chi^2$-minimisation procedure using a 
Levenburg-Marquardt algorithm \citep{press89}, 
 as implemented in the package {\sc sfit} \citep{jeffery98b},  
was used to solve for \teff, \logg, \nhe/\nh, and the velocity 
shift $v$. Being the velocity-shifted mean of several spectra having different heliocentric velocities, 
the latter had no physical meaning for this spectrum.

Re-normalisation was iterated to convergence with the final solution, 
with convergence represented by changes in successive solutions
becoming smaller than one tenth the formal errors. 
The final re-normalised spectrum and best-fit theoretical spectrum are shown in 
Fig.~\ref{f:alfosc}, with principal H, He{\sc i} and He{\sc ii} lines identified,
these providing the primary constraints on \teff, \logg, and \nhe/\nh.
The adopted continuum windows are also identified in Fig.~\ref{f:alfosc}. 

Using a grid of models with solar metallicity composition ({\bf p00})
and (H:He) number fractions (0.9:0.1), (0.7:0.3), and (0.5:0.5), we obtained  
$T_{\rm eff}=39\,100\pm250$\,K,  $\lgcs=6.02\pm0.12$ and
$\nhe = 0.21\pm0.05$   ($\log \nhe/\nh = -0.57\pm0.01$) 
The errors are formal statistical errors in the fit. 

Changing the initial values in the multi-parameter fit (e.g. from below the
final solution to above the final solution) had no effect on the result. 
Small changes in the definition of
continuum windows led to systematic errors of 
$\delta T_{\rm eff}  \approx \pm100$\,K, $\delta \logg \approx \pm0.1$ and 
$\delta \log  \nhe/\nh \approx \pm  0.02$.
Reducing the width of the Gaussian used to smooth the continuum to 50\AA\ 
(from 200\AA)  provided some benefits for continuum fitting at the
possible expense of allowing the fit to lead the final solution, in this case 
by  $\delta T_{\rm eff}  = -190$\,K, $\delta \logg = +0.15$ and 
$\delta \log  \nhe/\nh = -0.03$.

The initial assumption of a solar-mixture of metals is inconsistent with the 
observations; some lines  appeared in either the observation or the model which do not 
appear in the other.  This inconsistency  was partially resolved following the 
fine-abundance analysis described in \S~\ref{s:abunds}  and Table\,\ref{t:abunds}
and the computation of a grid of models with fixed helium abundance (\nh:\nhe=80:20)
approximating the star's surface composition  ({\bf h80he20\_uvo0825}). 
Repeating  $\chi^2$-minimization and re-normalisation  led to a final solution 
with $\teff= 38\,900\pm270$\,K and $\lgcs=5.97\pm0.11$ (formal errors). 
 
These results give  $\delta T_{\rm eff} = 2500$\,K and $\delta \logg = 0.35$ higher than, and  $ \log \nhe/\nh$ 
similar to those given by \citet{vennes11};  \citet{nemeth12} obtained a surface gravity and helium abundance 
similar to the new measurement, but with \teff\ more similar to that of \citet{vennes11}. 
The \citet{vennes11} result was based on non-LTE models containing hydrogen and helium only, with 
no allowance for back-warming due to opacity from other elements, or to the 
blanketing effects due to ultraviolet metal lines, which we
have already shown to be significant (Fig.~\ref{f:iue}). \citet{nemeth12} included more species in their 
customized  non-LTE model atmosphere calculation, but were only able to measure the 
abundance of one element, nitrogen, with upper limits for carbon and oxygen.

In order to assess the impact of the LTE approximation, a non-LTE model atmosphere 
was obtained using the T\"ubingen Model Atmosphere Package ({\sc tmap}) at the 
German Astrophysical Virtual Observatory {\bf \citep{werner12,ringat12}\footnote{http://dc.g-vo.org/theossa}}. 
The parameters were $\teff=40\,000$\,K, $\lgcs=6.0$,  
and composition approximately matched to that of the {\bf h80he20\_uvo0825} grid. 
The model  included ions of hydrogen (2 ions), helium (3), carbon (5), nitrogen (5), oxygen (5), neon (5), 
sodium (5) and magnesium (5),  represented by a total of 1435 LTE levels and 486 non-LTE levels. 
The formal solution for the blue-optical spectrum was convolved with a 1\AA\ Gaussian and 
rebinned at 0.4\AA, in order to match as closely as possible the {\it NOT/ALFOSC} spectrum.
 Using {\sc sfit} and  the {\bf h80he20\_uvo0825} grid, a best-fit match to the {\sc tmap} model
was sought in the same way as before, yielding $\teff=40\,740\pm130$\,K and $\lgcs=6.18\pm0.06$.  
This is only an indication of the systematic errors; whilst the {\sc tmap} model drops the
approximation of LTE, it also omits a substantial amount of line opacity in the ultraviolet, 
with consequences for the temperature structure of the photosphere. 


\section{Abundance Analysis}
\label{s:abunds}

\begin{figure}
\epsfig{file=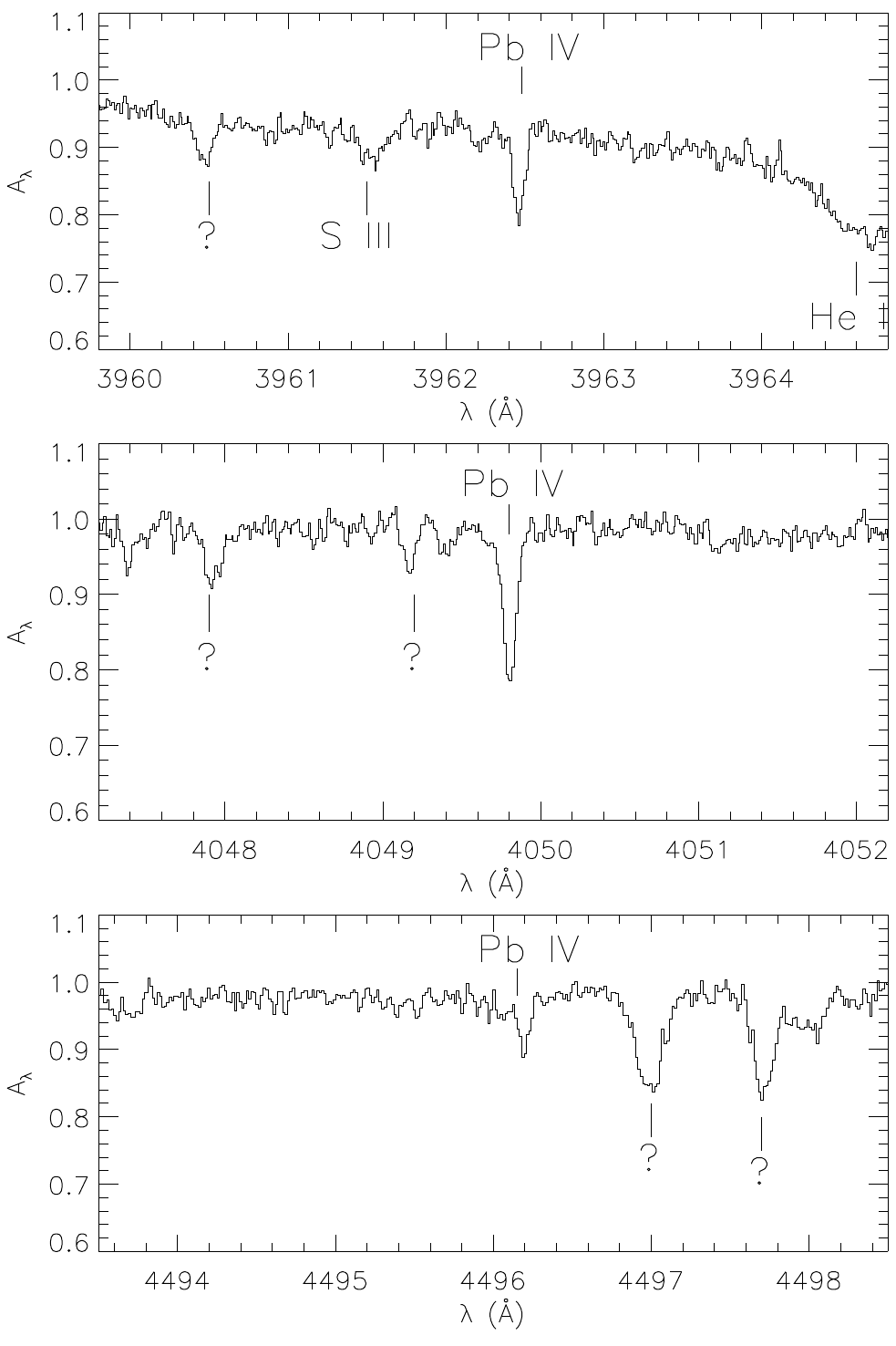,width=80mm}
\caption{Absorption lines dues to triply-ionized lead in the {\it Subaru HDS} spectrum
of \ledz. The spectrum has been shifted to the laboratory rest-frame. Identified  lines are marked by ion; unidentified lines are marked `?'. }
\label{f:pbiv}
\end{figure}

A first inspection of the {\it Subaru/HDS}  spectrum of \ledz\ revealed  the presence of
several strong lines due to triply ionized lead (Fig.\,\ref{f:pbiv}), 
one of the defining characteristics of this star and one demanding an urgent and 
detailed abundance analysis.  

A major difficulty with the analysis of \'echelle spectra of high-gravity 
hot stars is the calibration of the flux, and particularly the definition of
the continuum, due to the extent of the broad Balmer and helium absorption
lines over entire and adjacent \'echelle orders. Thus it is risky to 
rely heavily on atmospheric parameters derived solely from such lines. 
However, in the current case, the {\it NOT/ALFOSC} spectrum provides 
a reference, and enables a robust approach to measuring the surface 
abundances in \ledz. 

The process essentially involves renormalizing the {\it Subaru/HDS} spectrum to 
the theoretical spectrum, using the previous fit and the  model to determine 
the broad hydrogen-Balmer and helium line profiles, but allowing the 
narrow-line absorption spectrum to be analyzed correctly. This is especially
important for our analysis software which relies on a precise definition of
the local continuum, and also  for treating narrow lines which occur 
in the wings of much stronger lines.  To confirm that the process works, 
the renormalized  spectrum was analyzed  for \teff, \logg\ and \nhe\  as before, 
giving results fully consistent with expectation.

\subsection{Line lists}

Having adopted a model atmosphere for the abundance analysis, the primary
requirement for the abundance analysis is atomic data. Since we are making the
assumption of LTE, the only data required for each line, are the
wavelength, oscillator-strength, collisional and radiative damping constants 
(van der Waal's damping is included but not important), and the excitation 
potential of the level from which the line originates. 
In common with previous studies of intermediate helium hot subdwarfs \citep{naslim11,naslim13}
this investigation encountered several lines rarely (if ever) seen in an 
astronomical context. Therefore considerable effort was made to check the
completeness and reliability of the linelist adopted. 

Construction of a line list to analyse  blue-optical spectra  ($\lambda 3600 - 5200$ \AA) 
of early-type stars has progressed over several decades, commencing with 
{\sc lte\_lines} \citep{jeffery91},which focused on assessed data for 
single-line studies.  This list was augmented  to include other lines present in the 
spectra of mid- to late-B stars in order to synthesise large regions of the optical-blue 
spectrum \citep{woolf02c}.  Many of these data were contained within the 
compilation distributed by \citet{kurucz_cd23}. More recently, 
new atomic data were computed for zirconium, yttrium, germanium and 
lead lines discovered  in helium-rich hot subdwarfs  \citep{naslim11,naslim13}. 
On computing a spectrum including all of these species to compare with 
the {\it Subaru/HDS}  data, it was clear that many observed lines had no counterpart 
in the model,  even after accounting for possible abundance excesses. 
Since \ledz\ is significantly hotter than most stars previously analyzed by us, 
on-line atomic databases were examined to assess whether other ions or more
recent data could make up the deficit. The NIST Atomic Spectra Database \citep{nist15}
was used extensively to identify possible candidates. Additional lines of 
Ca$^{++}$ and S$^{++}$ were identified in this way, and associated atomic were located
in the  Vienna Atomic Line Database (VALD) \citep{vald95,vald15}. 

\paragraph*{Unidentified lines.} 
Efforts to identify additional lines included collating data for  double and triply
ionized species from VALD, 
setting artificially high abundances for these species and comparing 
the resultant spectra with observation. 
With over 150 lines still unidentified, other  possibilities had to be considered: \\
{\it 1. Artefacts.} Whilst instrumental artefacts or noise in the raw data can account for some very
weak features, the vast majority  are true absorption lines with equivalent widths 
above the detection threshold of 5\,m\AA. \\
{\it 2. Second Star.} There is no evidence that the spectrum of a second star is present. 
Wherever a line was eventually identified it occurred at the correct wavelength (or radial velocity).
Strong unidentified lines were checked against spectra of late-type stars, and 
chemically-peculiar A stars, with no matches found. Any such match would 
have been surprising since there is no evidence of a flux-excess at visual or infra-red 
wavelengths from a cooler secondary. \\
{\it 3. Completeness.} The line lists  are almost certainly incomplete for stars of this effective 
temperature.  It is likely that many arise from ions which are not represented 
(cf. Table \ref{t:lines}),  but it is also possible that  the weakest lines are due to 
absorption from  high excitation states in ions already represented. 

\begin{table*}
\centering
\caption{Elemental abundances for \ledz\ and related stars in the form  $\log \epsilon_i = \log n_i + c$ (see text). Measurement errors are shown in parenthesis. The absence of a reported measurement is indicated by ``$-$''.   }
\label{t:abunds}
\setlength{\tabcolsep}{2pt}
\begin{tabular}{@{\extracolsep{0pt}}p{27mm}l lllll lllll l}
\hline
Star & H & He & C & N & O & Ne & Mg & Al & Si & S & Cl  \\
\hline
\ledz$^a$               &       &      &  $<6.5$ & 8.04(24) & 7.43(07) & 7.48(25) & 6.25(11) & 6.47(07) & 6.26(21) & 7.61(18) & 6.34(11)   \\
\ledz$^b$               & 11.8$^x$ & 11.2$^x$& [6.5] & 8.07    &  7.38   &    7.85 &  6.27  & 6.26 &  6.80 &  7.71  & 6.17  \\
{\bf h80he20\_uvo0825}$^c$ &  11.8 & 11.2 &   6.0 & 8.0 & 7.2 & 7.3 & 7.0 & 5.9 & 6.9 & 6.5 & [5.0]  \\[1mm]
\crimson $^{1}$ & 11.83 & $11.23(05)$ & $8.04(22)$ & $8.02(20)$  & $7.60(17)$ & $<7.6$ & $6.85(10)$ & $-$ & $6.32(12)$ & $-$ & $-$  \\
HE\,1256$-$2738$^{2}$ &   11.45 & 11.44 & $8.90(54)$ & $8.14(62)$ & $8.08(10)$ &$<7.1$ & $<6.5$  & $-$ & $6.19(10)$ & $<6.5$ & $-$  \\
HE\,2359$-$2844$^{2}$ &   11.58 & 11.38 & $8.51(29)$ & $8.00(57) $ & $7.81(16)$ &$<6.9$ & $7.6(1)$ &$-$  &$5.73(13)$ & $<6.3$ & $-$  \\[1mm]
JL\,87$^{3}$       &    $11.62(07)$ & $11.26(18)$ & $8.83(04)$ & $8.77(23)$  & $8.60(23)$ &$8.31(57)$ & $7.36(33)$& $-$ & $7.22(27)$  &$6.88(1.42)$  & $-$   \\
PG\,0909$+$276$^{4,5}$ &    & 11.15(10) & 8.63(35) & 8.00(23) & $<7.50$ & $<7.87$ & $-$ & $<6.25$ & $5.80(10)$ &  8.26(53) & $-$   \\
UVO\,0512$-$08$^{4,5}$ &   & 11.23(10) & 8.59(20) & 7.94(21) & $<7.75$ & $-$ & $-$ & $<6.25$ & 5.76 & 8.14(49) & $-$   \\[1mm]
cool sdB$^{5,d}$       &      &  ~9.24(54) & 6.99(47) & 7.68(41) & 7.88(26) & $-$ & 6.54(26) & 5.70(18) & 6.79(37) & 6.51(21) & $-$ \\
warm sdB$^{5,e}$   &     & 10.15(76) & 7.73(70) & 7.42(27) & 7.67(51) & 7.27(67) & 7.17(29)& 6.2 & 6.02(55) & 7.18(56) & $-$ \\
Feige\,66$^{6}$   & & 10.4 &  6.79(30) & 7.65(15)  & $-$ & $-$ & $-$ & $<3.5$ & $<2.0$ & 7.69(46) & $-$   \\[1mm]
Sun$^{7}$          & 12.00     &[10.93]& 8.43 & 7.83  &8.69  & [7.93] &  7.60  & 6.45 & 7.51 & 7.12   & 5.50    \\[3mm]
\hline
Star  & Ar & Ca & Ti & V & Fe  & Ge & Sr & Y & Zr & Pb      \\
\hline
\ledz$^a$          & $<8.3$  & 8.31(21) & 7.37(34) & 7.51(25) & $< 7.0$ & 6.24(06)  & $-$ & $5.37(09)$ & $<5.3$ & 5.49(18) \\
\ledz$^b$        & [6.6] & 8.14 & 7.44  & 7.46 & 7.30 & 6.16 & [6.9] & 5.37 &  [2.6] & 5.46 \\
{\bf h80he20\_uvo0825}$^c$ & 5.9 &  8.6 & [7.6] & [6.5] & 7.5 & [3.7] & [2.9] & [2.2] & [2.6] &  [1.9]  \\[1mm]
\crimson $^{1}$ & $-$ & $-$ & $<6.0$ & $<6.5$ & $<6.8$&  6.28(12) & 6.96(15) & 6.16(10) & 6.53(24) & $-$\\
HE\,1256$-$2738$^{2}$ & $-$  & $-$ & $-$ & $-$ & $-$ & $-$ & $-$ & $-$ & $-$ & 6.39(23)\\
HE\,2359$-$2844$^{2}$ & $-$ & $-$ & $-$ & $-$ & $-$ & $-$ & $-$ & 6.61(15) & 6.47(15) & 5.64(16) \\[1mm]
JL\,87$^{3}$   &  $-$ & $-$ & $-$ & $-$ &  $-$    & $-$ & $-$ & $-$ & $-$ & $-$  \\
PG\,0909$+$276$^{4,5}$ & 8.68(15) & 7.81(35) &  7.97(20) &   8.10(26)$^y$ & $<7.87$ & $-$ & $-$ &$-$ &$-$ &$-$\\
UVO\,0512$-$08$^{4,5}$ &  9.90 &  8.10(24) & 8.06(33) & 7.36(22)$^y$  & $<7.81$ & $-$ & $-$ &$-$ &$-$ &$-$\\[1mm]
cool sdB$^{5,d}$      & 6.78(21)  & $-$ & 6.30(35) & 7.10(36) & 7.58(20) & $-$ & $-$ &$-$ &$-$ &$-$ \\
warm sdB$^{5,e}$    & 7.89(17) & 7.98(25) & 7.04(36) & 7.78(20) & 7.46(24) &  $-$ & $-$ &$-$ &$-$ &$-$ \\
Feige\,66$^{6}$   & 7.86(24) &  8.09(20) & 6.96(22) & 6.37(22) & 6.46(17) & 5.21(05) & $-$ & $-$ & $-$ & 4.7  \\[1mm]
Sun$^{7}$  & [6.40] & 6.34 & 4.95 & 3.93 & 7.50         & 3.65 & 2.87 &  2.21 & 2.58 &  1.75 \\
\hline
\end{tabular}\\
\parbox{170mm}{
Notes:\\
$a$. abundances from line equivalent widths.\\
$b$. abundances from spectral synthesis $\chi^2$ minimization;  values in square brackets were fixed (i.e. no solution was obtained). \\
$c$.  model atmosphere input abundances as used in  Fig.~\ref{f:iue} and \ref{f:alfosc}; values in square brackets refer to species for which continuous opacities were not included. \\
$d$. $25\leq \teff/{\rm kK}\leq 27$. \\
$e$. $35\leq \teff/{\rm kK}\leq 40$ excluding PG\,0909$+$276 and UVO\,0512$-$08.\\
$x$. from {\it NOT/ALFOSC} spectroscopy \\
$y$. from \ion{V}{iv}.
}\\[1mm]
\parbox{170mm}{
References: 
1. \citet{naslim11}, 
2. \citet{naslim13},
3. \citet{ahmad07},
4. \citet{edelmann03},
5. \citet{geier13},
6. \citet{otoole06}, 
7. \citet{asplund09}; photospheric except helium (helio-seismic), neon and argon (coronal). 
}
\end{table*}

\subsection{Methods}

The process of measuring abundances was carried out in two ways. The first approach
was to compute a best-fit theoretical spectrum in which only elemental 
abundances were allowed to vary. A best-fit solution was obtained by $\chi^2$-minimization using a 
Levenburg-Marquardt method \citep{press89}, \citep[\sc sfit: \rm][]{jeffery98b}  
The result is shown in Appendix\,\ref{s:lines}. 

A solution was sought for each element separately. Starting with an 
 estimate of abundance (e.g. three times solar) a solution was sought and noted.  This solution 
was then fixed in  the starting mixture used for investigating subsequent elements. 
The final solution was obtained after iteration to allow for blended lines.
Tests with different start estimates were made to insure against finding false local minima. 

The second approach was to measure equivalent 
widths and compute abundances for selected lines from an individual curve-of-growth 
computed for each line. The micro-turbulent velocity was verified by the same means;
the mean  abundance obtained from 21 \ion{N}{ii} lines with $5 < \ew/{\rm m\AA} < 30$ 
drops by less than 0.05 dex as \vt\ is increased from 2 to 20 \kms; at this \teff, the
lines appear insensitive to small values of \vt, probably because the latter are small relative 
to the thermal broadening at this effective temperature. We adopted $\vt=2\kms$ in line with \citet{geier13}.  
The line-by-line abundance measurements for unblended lines with 
atomic data available are shown in Appendix\,\ref{s:lines}.

The principal sources of error in the abundances arise from the atomic data, 
which need to be consistent across all ions 
and multiplets but which frequently are not (cf. \ion{S}{iii}, Appendix\,A: Fig. A.2),  
and from line misidentifications and blends. $\chi^2$-minimization
uses all lines for which atomic data are available, whether present, blended
or hidden in noise. The equivalent width approach uses primarily unblended lines with $\ew > 5 {\rm m\AA}$. 
This systematic difference in the line samples, coupled with the error types identified, contributes 
to small differences provided by the two methods. Where significant, the equivalent-width 
measurements are preferred. 

From the width of the lead absorption lines, the projected rotation velocity is small. 
The overall quality of the spectrum prevents us from setting a limit any stronger than $v \sin i < 10\kms$

\subsection{Abundances}

Abundances derived for each element identified in the spectrum of \ledz\ are 
shown in Table\,\ref{t:abunds}, together with those adopted for the model atmosphere and data for comparable stars. 
Abundances are given in the form $\log \epsilon_i = \log n_i + c$ where $\log \Sigma_i a_i \epsilon_i = 12.15$,   $a_i$ are 
atomic weights, and $c = \log \Sigma \epsilon_i$. This conserves values of $\epsilon_i$ for elements whose abundances 
do not change, even when the mean atomic mass of a mixture changes substantially, and conforms to the convention that $\log \epsilon_{\rm H} \equiv 12$
for the Sun and other hydrogen-normal stars.

Errors are based on the standard deviation of the line abundances about the mean or, in the case of a single
representative line, on the estimated error ($\pm 2$\,m\AA) in the equivalent width measurement  (Table\,\ref{t:lines}), 
and an assumed error of $\pm10\%$ in the transition probability. 
In the case of \ledz\, upper limits were obtained by computing the abundance required to give a line with an 
equivalent width of 5\,m\AA, the detection threshold in the {\it Subaru/HDS} spectrum. 

Carbon is not detected in the spectrum of \ledz. The strongest predicted line \lineA{C}{iii}{4070.3} is not present. 
With abundance $\log \epsilon_{\rm C}=6.5$ (2 dex below solar) this line would be predicted to have an equivalent 
width 5.6\,m\AA.  Similarly, iron is at the boundary of detectability, although a solution is obtained with {\sc sfit}. 
Notably, germanium, yttrium and lead are measured with abundances 2.5 -- 4 dex above solar. 

A few discrepancies between the equivalent-width,  $\chi^2$-minimization  and input model 
abundances persist (Table\,\ref{t:abunds}). The model inputs are  restricted, with elements heavier than 
oxygen being defined in groups rather than individually and their abundances being chosen as a compromise 
amongst several elements. Reasons for differences between line-by-line and  $\chi^2$ measurements 
have been discussed above.

\paragraph*{Heavy-metal subdwarfs.}
For comparison with \ledz\, abundances for the zirconium-rich \crimson\ and two intermediate He-sds 
with extreme overabundances of lead, \rooster\ and \kraft,  are shown in Table\,\ref{t:abunds}. 

\paragraph*{Intermediate-helium subdwarfs.}
Apart from hydrogen and helium, the prototype intermediate helium subdwarf JL\,87 appears to have a 
roughly solar-like surface composition \citet{ahmad07}, but no elements heavier than sulphur were measured. 
\citet{edelmann03} identified two intermediate helium-rich subdwarfs, PG\,0909$+$276 and  UVO\,0512$-$08, 
with extreme overabundances of some iron-group elements, including scandium, titanium, vanadium, 
manganese and nickel, but {\it not} iron (Table\,\ref{t:abunds}). Eight other intermediate helium subdwarfs for which 
abundances have been measured show no detectable excesses
\citep{naslim10,naslim12,naslim13} although, in many cases, even substantial excesses could not be detected 
with available spectra. 
Abundance analyses for other intermediate helium stars having $0.1 < \nhe < 0.9$ and no measured excess in any of Ge, Sr, Y, Zr, or Pb 
include BPS CS 22956--0094 \citep{naslim10}, CPD--20$^{\circ}$1123 \citep{naslim12}, 
HE\,0111$-$1526, HE\,1135$-$1134, HE\,1238$-$1745, HE\,1258$+$0113, HE\,1310$-$2733, and 
HE\,2218$-$2026 \citep{naslim13}. 

\paragraph*{Normal subdwarfs.}
\citet{geier13} published surface abundances for a sample of 106 hot subdwarfs, the large majority of which are helium poor. 
He demonstrated that some elements show significant trends with effective temperature. Table\,\ref{t:abunds} includes
mean abundances in two ranges of \teff, representing cool and warm sdB stars, between which there is a smooth trend 
across the entire sdB temperature range (Fig.\,\ref{f:abunds}). The warm group corresponds to the temperature 
range of the heavy-metal subdwarfs. Both extremes demonstrate significantly
sub-solar abundances of light elements, notably helium, oxygen, magnesium, aluminium, and silicon, roughly solar abundances of 
nitrogen and iron, and super-solar abundances of calcium, titanium and vanadium. Carbon is under-abundant at
effective temperatures below 35\,kK, above which the mean value approaches solar.
\citet{otoole06} were able to explore additional elements using {\it HST} spectroscopy of five hot subdwarfs, 
establishing, further to the above, super-solar abundances of lead and, in some cases, germanium. Of these, 
Table\,\ref{t:abunds} includes the results for Feige\,66.


\begin{figure*}
\epsfig{file=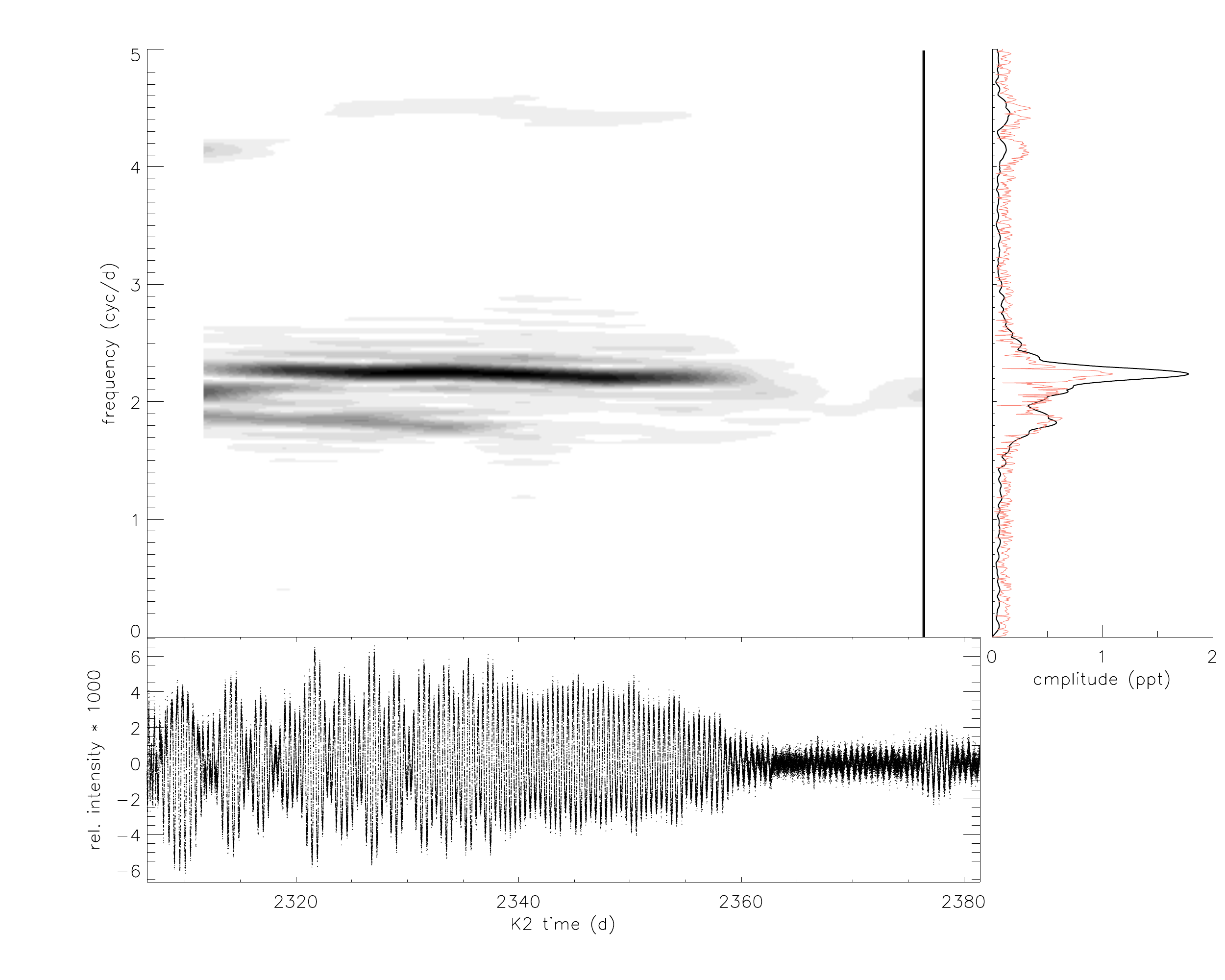,width=175mm,angle=0}
\caption{Main panel: the sliding amplitude spectrum of the {\it K2}  light curve of \ledz\ shown as greyscale, based on 
 data blocks of duration $\Delta T=10$\,d sampled approximately every 3.7\,d.  The frequency resolution ($1/\Delta T$) 
is hence $\pm0.1\perday$.  The light curve is reproduced at the same horizontal scale in the panel beneath. The panel on the right shows, in bold, 
the time-averaged amplitude spectrum,  and in light red, the amplitude spectrum of the entire dataset. 
This figure demonstrates the presence of  power around two principal frequencies at  $2.23\pm0.01$ and $1.82\pm0.04 \perday$ 
and at higher frequencies. } 
\label{f:slide}
\end{figure*}

\begin{table}
\caption{ Principal frequencies and semi-amplitudes (ppt = parts per thousand)  
measured from the entire {\it K2} dataset and for subsets of the {\it K2} data for \ledz.  }
\label{t:freqs}
\setlength{\tabcolsep}{4pt}
\begin{tabular}{rr rrr rrr}
\hline
{\it K2}    $t_0$- $t_1$ & $\delta f$ & $f_1$ & $\sim2f_1$ & $f_2$ & $a_{f1}$ & $a_{2f1}$ & $a_2$ \\
  d    &  \perday & \perday & \perday &  \perday & ppt & ppt & ppt \\
\hline
2306.4--81.4	&0.013	&2.237 &4.497 &1.861 &1.19 &0.12 &0.40  \\[2mm]
2306.6--14.1	&0.133	&2.077 &4.126 &          &2.36 &0.92 & \\
2314.1--21.6	&0.133	&2.278 &3.930 &1.822 &2.52 &0.26 &1.36 \\
2321.6--29.0	&0.133	&2.248 &4.477 &1.828 &3.36 &0.37 &1.77\\
2329.0--36.5	&0.135	&2.252 &4.494 &1.778 &3.40 &0.43 &1.30\\
2336.5--44.0	&0.133	&2.228 &4.448 &1.905 &3.42 &0.42 &0.62\\
2344.0--51.5	&0.133	&2.200 &4.391 &1.860 &3.22 &0.40 &0.65\\
2351.5--59.0	&0.133	&2.197 &4.407 &          &2.17 &0.21 & \\
2359.0--66.5	&0.133	&2.054 &3.919 &          &0.40 &0.08 & \\
2366.5--74.0	&0.133	&1.909 &3.824 &1.909 &0.35 &0.05 &0.35 \\
2374.0--81.4	&0.133   &2.073 &4.257 &          &0.67 &0.05 & \\[2mm]
2306.6--66.6	&0.017	&2.236 &4.497 &1.858 &1.47 &0.15 & 0.52  \\
\hline
\end{tabular}
\end{table}

\section{Light Curve Analysis}
\label{s:light}

A first examination of the light curve is confusing. The first part shown in Fig.~\ref{f:k2} suggests
a period around 0.5\,d modulated by a signal having a beat period of  about 5 days. The 0.5\,d
period appears to be present throughout most of the {\it K2} observations (Fig.~\ref{f:k2}), 
with significant amplitude modulation at early times,  little modulation at middle times, and
decaying to negligible amplitude at late times. In order to interpret the light curve, it is
first necessary to establish its period content. 

An inspection of the amplitude spectrum obtained from the discrete Fourier transform  is 
equally  perplexing. The amplitude spectrum for the entire {\it K2} light curve shows a
group of peaks with frequencies $f$ around $2 {\rm d^{-1}}$,  a significant peak at  
$f_1\approx 2.2 {\rm d^{-1}}$, and some power at  $\approx2f_1$. The amplitude of the 
signal at $f_1$ is, however, substantially weaker than that seen in a large part of the  light curve. 
Dividing the light curve into segments shows that there is considerable variation in the
amplitude spectrum over time (Fig\ref{f:k2}). Two primary reasons suggest themselves: i) the signal
is non-uniform, varying in both amplitude and frequency and ii) there are multiple signals
present and unresolved in the relatively short duration of the {\it K2} campaign. Other 
interpretations are also be possible. 

The light curve was investigated using a sliding Fourier transform, an example of
which is shown in Fig.~\ref{f:slide}. The choice of the duration of dataset in each element
of the sliding transform is a compromise between temporal and frequency resolution;
samples of duration 10\,d giving a frequency resolution of $\approx 0.1\perday$ were found
to give the most coherent picture of the current data. The light curve was also investigated by dividing into
ten discrete subsamples, computing the amplitude spectrum, and measuring the
frequency and amplitude of the principal peaks for each sample, as
well as for the entire spectrum and various other subsamples  (Table~\ref{t:freqs}).

The picture that emerges from this analysis is that a dominant oscillation with $f_1\approx2.24\pm0.01\perday$
and semi-amplitude $a_1\approx3$ parts per thousand (ppt)
persists throughout most of the {\it K2} light curve. At early times, an additional oscillation
with  $f_2\approx1.8 \perday$ is observed. The relative amplitude of the two signals is not well determined; 
typically $a_1/a_2\approx2 - 5$.  In Fig.\ref{f:slide}, $f_1$ appears to vary, but only by an amount 
which is less than the frequency resolution  ($\pm 0.1\perday$).  At the very start of the {\it K2} campaign 
$f_1$ appears to be significantly shorter at $\approx2.08\pm0.13\perday$. 
Such behaviour might be anticipated if, for example, the frequency content of the power 
spectrum is not fully resolved. The frequency resolution of the entire dataset is $\delta f \approx0.013\perday$.
Signals which can only be identified for a part of the time series are less well resolved ($\delta f \approx0.13\perday$).
For signals with frequencies $\approx 2\perday$, both resolutions are low; the data thus suggest the presence of multiple 
unresolved closely-spaced frequencies.  

 
On the basis of available evidence, it appears that the light variations in \ledz\ can be
interpreted in terms of a multi-periodic signal with multiple unresolved 
frequencies around 1.8 and 2.2 \perday  (periods 10.8 and 13.2\,h). The usual explanation for such a signal is the presence of 
global oscillations occurring simultaneously in a number of modes. For a star of the dimensions of \ledz, the fundamental radial
mode would have a period of $\approx300$\,s \citep{jeffery16a}; if the oscillations arise from
\ledz, they must be associated with gravity modes of extremely high radial order.  
There is currently no known mechanism which could excite such modes
in \ledz, which is an even more extreme example than \crimson\, in which g-modes with periods of
$\approx 1800$\,s have been observed \citep{ahmad05,green11}. 

Other interpretations for light variations with the observed period and amplitude must be considered. 
Reflection from a  low-mass companion (a planet perhaps) with an inhomogeneous 
surface might suffice, especially if rotating asynchronously.  However, assuming an albedo of unity, 
a Jupiter mass planet in a 12 hour orbit would  reflect $\approx 0.6$ ppt of the parent star's light.
A 0.3\Msolar dwarf {\it could} reflect $\approx 4$ ppt of the parent starlight, but would require an 
orbital velocity $\approx90\kms$ from the primary. 
Reducing the inclination to  match the non-detection of   orbital motion (\S\,\ref{s:not_obs})  
would reduce the apparent light variation below that detected.
 Companions of sufficiently low mass ($\leq0.01\Msolar$) and  large radius ($\geq0.25\Rsolar$) 
 to satisfy the radial velocity {\it and}  light curve constraints are otherwise difficult to identify 
in astronomical terms.  Further difficulties with such a solution include the apparent drift in frequency, 
and the absence  of evidence of any infrared excess (\S\,\ref{s:fluxes}). 
Geometrical effects due to tidal distortion by a  massive companion can be ruled out on similar grounds. 
The possibility of light modulation by  magnetically supported  surface spots was discussed in the context of \crimson\ 
by \citet{ahmad05},  discounted by \citet{green11}, and eliminated by \citet{randall15}. 
However,  it remains necessary to consider a differentially rotating surface with migrating spots as a possible
light curve driver. Combining the upper limit to $\vrot < 10\kms$ (\S\,\ref{s:abunds}) with a typical 
hot subdwarf  radius $\approx 0.12 \Rsolar$ gives a miniumum rotation period of $\approx14.6\,$h,  
which is almost compatible with the observed periods of 11-13\,h. 

Due to the large size of the detector pixels ($\approx4^2$ arcsec), a problem frequently encountered 
in the analysis of both {\it Kepler} and {\it K2} data is
contamination by  light variation of a nearby star \citep[cf.][]{silvotti14} or, less frequently, 
charge leakage from a bright star occupying the same CCD column. The 2MASS catalogue shows 5 stars within
1 arcmin of \ledz, all having $m_{\rm J}>15.3$, the closest having $m_{\rm J}=16.3$ is 20 arcsec distant
(\ledz\ has $m_{\rm J}=12.4$).  None are bright enough to contaminate the target in the manner observed. 
We found no evidence for any other potential source of contamination
despite investigating i) the behaviour of individal sky pixels in the mask, ii)  the  light curves for all
stars brighter than $14^{\rm th}$ mag. within 10 arcmin, and iii) potential contaminants in the full-frame image 
for \ledz. Furtermore, there is no evidence for an unresolved companion in the flux distribution (\S\,\ref{s:fluxes}).

The resemblance of the \ledz\ light curve to those of slowly pulsating B 
stars, especially KIC\,11293898 \citep[cf.][]{mcnamara12}, is remarkable and suggestive. 
From inspection of the complete light curve obtained over the entire {\it Kepler} mission, KIC\,11293898 shows 
an amplitude similar to that of \ledz,  with a rich cluster of well-resolved g-modes at frequencies around 
2.4  \perday, with harmonics at twice this value (and beyond). \ledz\ might be expected to show similar behaviour 
if observed continuously  for 3 years. How this can be reconciled with its spectroscopic properties remains a mystery.


\begin{figure}
\epsfig{file=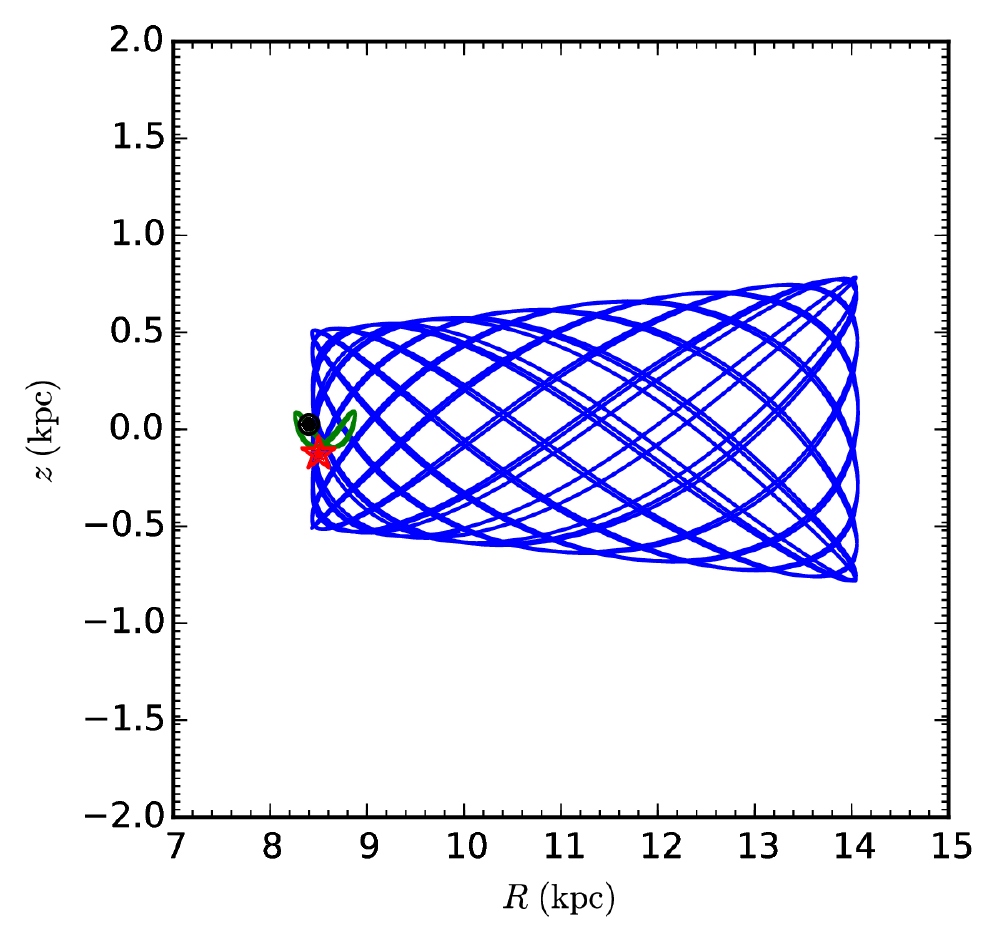,width=80mm,angle=0}
\epsfig{file=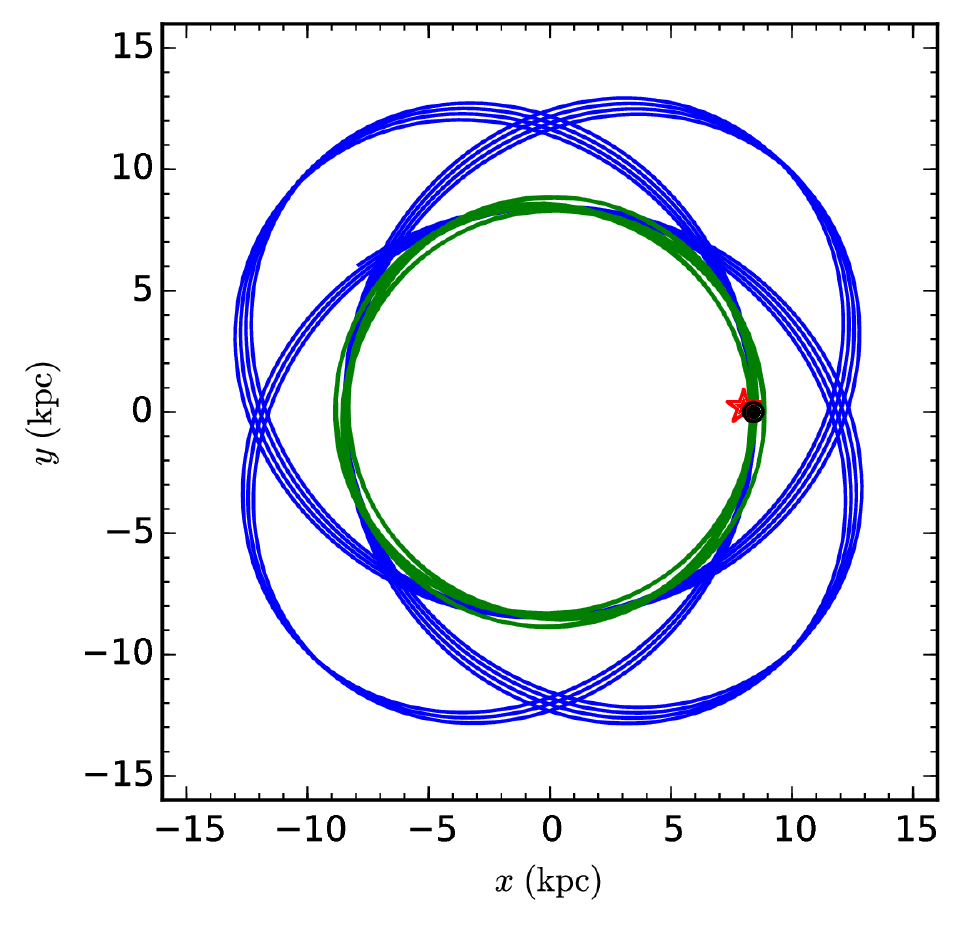,width=80mm,angle=0}
\caption{The current position (red star) and galactic orbit (blue) of \ledz\ projected forward over 3 Gyr. 
Top: meridional section ($R-z$). Bottom: projection onto the Galactic plane ($x-y$). 
The current position (black circle) and projected orbit (green) of the Sun are also shown. 
} 
\label{f:orbit}
\end{figure}

\begin{table}
\caption{Kinematical properties for \ledz.}
\label{t:kin}
\begin{tabular}{ll}
distance & $d = 257^{+37}_{- 29}\,{\rm pc}$ \\[1mm]
space motion &$U=-14\pm 2 \kms$\\
&$V=+303\pm 18\kms$ \\
&$W=-32\pm 2\kms$ \\[1mm]
apocenter radius &$R_{\rm a} = 14.3\pm0.4\,{\rm kpc}$ \\
eccentricity &$e = 0.26\pm 0.01$\\
galactic rest frame velocity &$v_{\rm grf} =+298\pm14 \kms$\\
$z$-component of angular mom$^{\rm m}$ &$J_z = 2502\pm 23\,{\rm kpc} \kms$\\
maximum height above plane &$z_{\rm max} = 0.80\pm0.04\,{\rm kpc}$\\
\end{tabular}
\end{table}

\section{Kinematics}

Evidence for peculiar galactic orbits amongst other intermediate helium- and heavy-metal-rich
subdwarfs \citep{randall15,martin16} and measurement of a large proper motion in \ledz\
\citep{hog00} raises the question of kinematics and the identity of the star's parent population.
 
Using the measured value for surface gravity and an assumed mass
typical for other subdwarfs ($0.50\pm0.1\Msolar$) leads to a stellar radius of 
$r=0.121^{+0.018}_{-0.014}\Rsolar$. The angular diameter then yields a distance
$d=257^{+37}_{- 29}\,{\rm pc}$.
Together with published proper motion measurements and the measurement of radial velocity from the {\it
Subaru/HDS} spectrum,  the galactic orbit can be computed \citep{martin16}. 
The components of space motion and Galactic orbital elements are given in Table~\ref{t:kin}.

With a Galactic rotation velocity of $V=+303\kms$, \ledz\ is rotating slightly faster than the Local
Standard of Rest (242\kms) and is typical of the thick disk population. Fig.~\ref{f:orbit} shows the
orbit morphology. The orbit extends  0.8\,kpc above the plane and has an orbital eccentricity of 0.26,
also indicative of thick disk membership.


\begin{figure*}
\epsfig{file=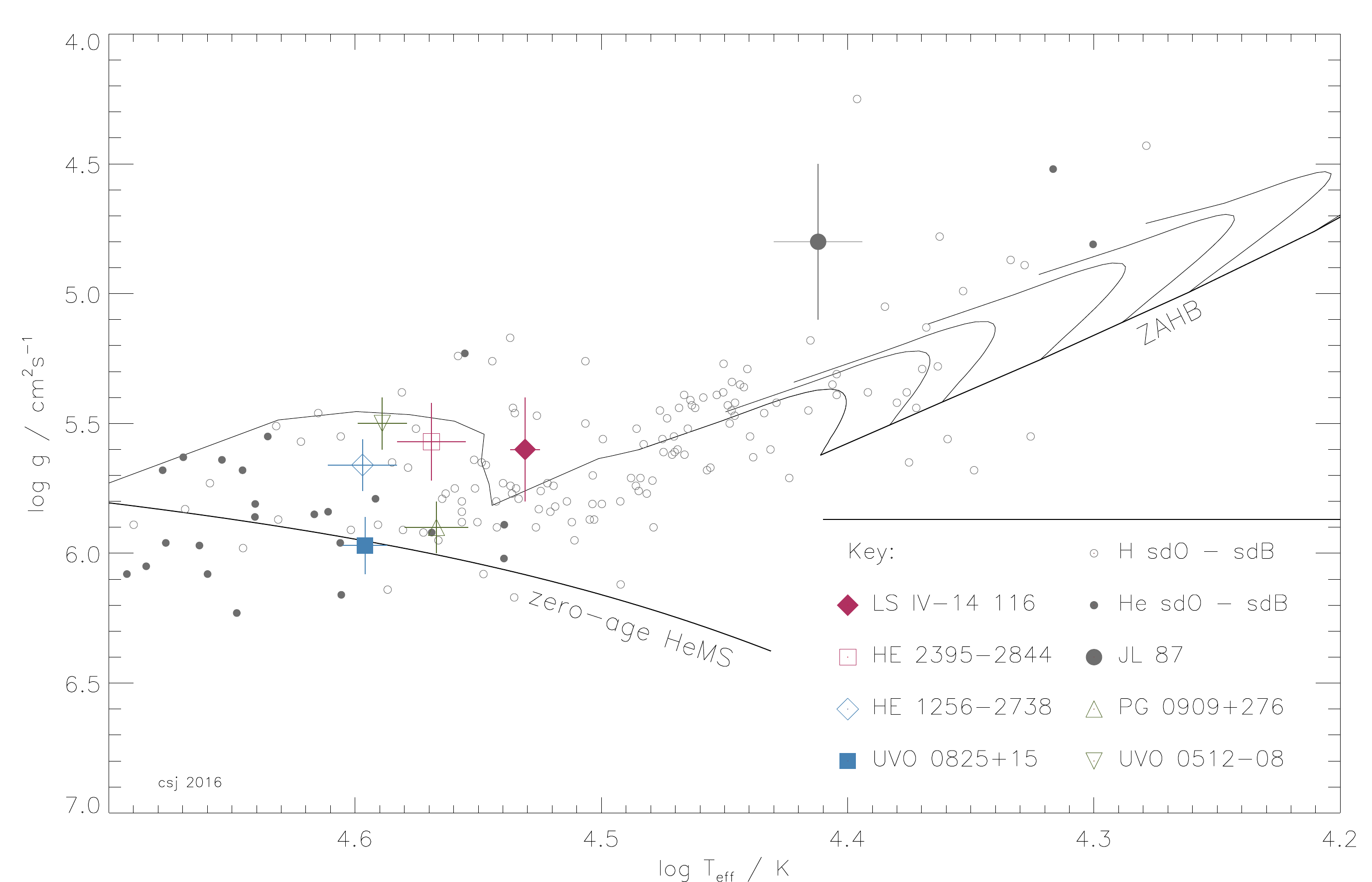,width=170mm,angle=0}
\caption{The distribution of  chemically-peculiar, helium-rich and normal hot subdwarfs with effective temperature and surface gravity. Solid lines show representative positions for the theoretical zero-age helium main-sequence (HeMS: $Z=0.02$) and models evolving from the zero-age  horizontal branch (ZAHB) to the end of core helium burning \citep[][; {\bf z22}: $M_{\rm c}=0.469, Y=0.288, {\rm [Fe/H]}=0.0$]{dorman93}. The post-HB  evolution of one model ($M=0.471\Msolar$) is also shown. The observed data are from this paper, \citet{edelmann03,ahmad07,naslim11,naslim13} and \citet{nemeth12}. } 
\label{f:tg}
\end{figure*}

\begin{figure*}
\epsfig{file=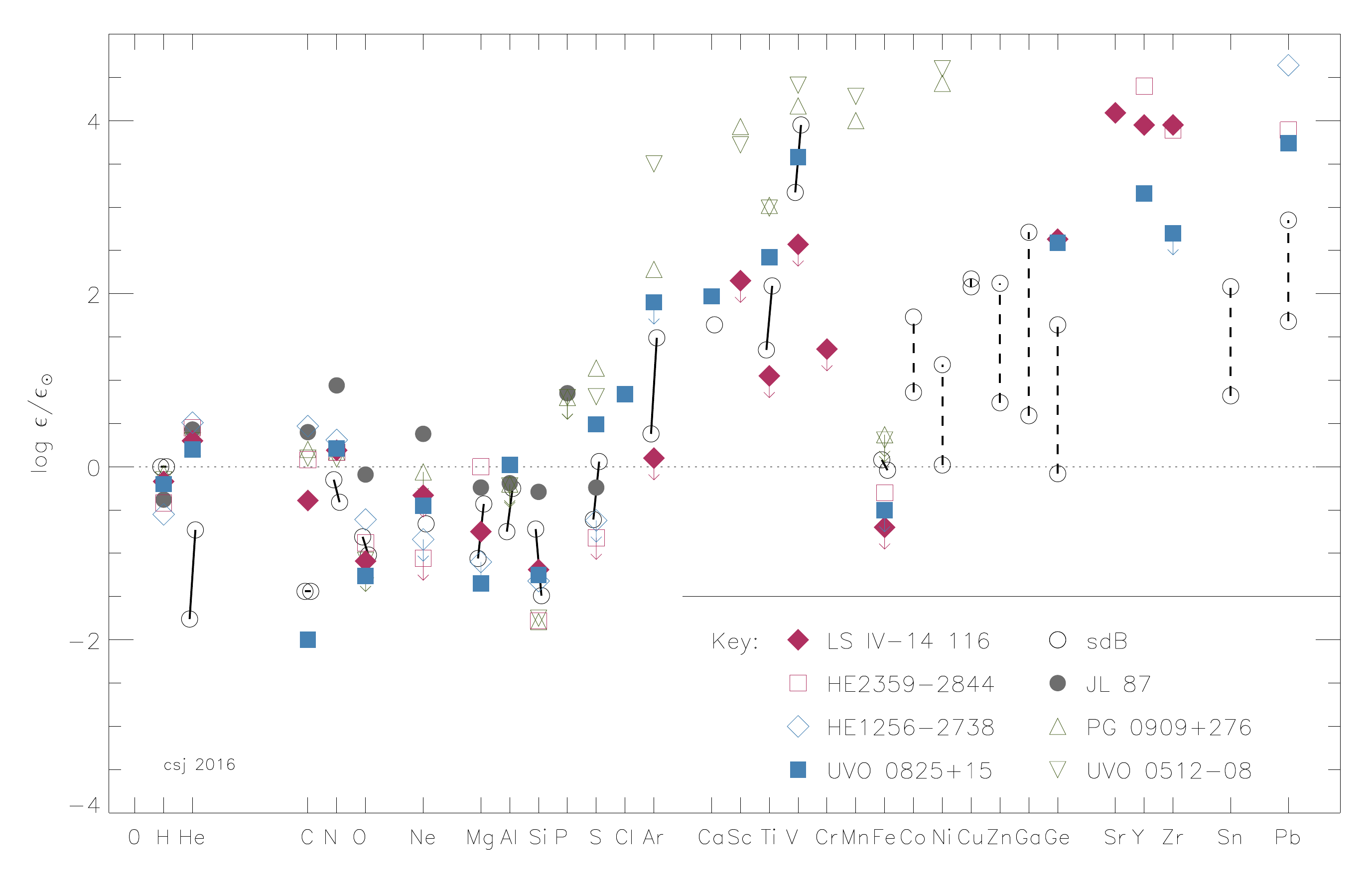,width=170mm,angle=0}
\caption{Surface abundances of super metal-rich hot subdwarfs, including the pulsating stars \crimson\ and \ledz\
\citep{edelmann03,naslim11,naslim13} and this paper. 
Abundances are shown relative to solar values \citep[dotted line: ]{asplund09}. Mean abundances and ranges for the helium-rich subdwarf JL\,87 \citep{ahmad07} and for normal 
subdwarfs are also shown.  The latter are shown by connected open circles as (i) $Z\leq26$ (solid lines): the  abundance distribution across 
the temperature range for sdBs as defined in Table\,\ref{t:abunds}, with cool and warm groups displaced left and right respectively \citep{geier13} 
and (ii) $Z\geq27$ (broken lines): the range of abundances measured for five normal sdBs from ultraviolet spectroscopy \citep{otoole06}.  }
\label{f:abunds}
\end{figure*}

\section{\ledz\ and other hot subdwarfs}

Figures\,\ref{f:tg} and \ref{f:abunds} place \ledz\ in context with other hot subdwarfs, including 
the ``normal'' helium-deficient   subdwarf B and subdwarf O stars, helium-rich hot subdwarfs, 
and a selection of chemically-peculiar 
intermediate helium-rich hot subdwarfs.  The latter, which include \ledz, 
lie in the range $30\,000 < T_{\rm eff}/{\rm K} <  40\,000$ and on or slightly above the helium
main sequence. They are sufficiently hot that they cannot be ``extended horizontal-branch'' stars
which retain a remnant hydrogen-rich envelope. Their position
on the helium main-sequence does suggest, but does not require, that they have helium-burning cores. 
If so, whatever hydrogen remains in their surface is sufficiently reduced in deeper layers that 
its opacity plays no role in the overall structure of the star. 
JL\,87 appears to represent another type of intermediate helium-rich subdwarf, being cooler
and having a near-solar distribution of light elements \citep{ahmad07}.  

\subsection{Chemistry}

The group including \ledz, \crimson, \rooster, and \kraft\ represent the true ``heavy-metal'' 
stars, having 3--4 dex excesses of some of germanium, strontium, yttrium, zirconium, and lead, 
this being at least 1 dex higher than seen in any other hot subdwarf. \quartz\ and \ethel\ show very high excesses of
iron-group elements, but no detection of the heavy metals. The simplest interpretation of
the peculiar chemistries of all six stars is that their radiation-dominated photospheres are 
modified by diffusive processes, including gravitational settling of heavy species and 
radiative levitation. The latter will concentrate 
atomic species into layers where their specific opacities are high, generally  
corresponding to the ionization temperatures of specific ions. If these layers also correspond to
layers in the photosphere where absorption lines form, then they will be observed with
large overabundances. The converse will also be true; ions of low specific opacity in the 
line-forming layer will migrate upwards or downwards and hence
appear under-abundant. 

Significantly, none of these  groups, {\it including} the normal sdB stars shows
 any surface excess of iron; it may be  depleted in some cases. As well as its heavy-metal overabundances, 
\ledz\ is  remarkable, and possibly unique, for its very low surface carbon abundance. 
It is tempting to use the iron abundance as an indicator of overall metallicity, 
and the carbon/nitrogen/oxygen ratio as evidence that the hydrogen-depleted surface 
is the  product of CNO-process hydrogen-burning. This is almost certainly misleading, 
since it  directly contradicts the argument that diffusion is responsible for the exotic 
chemistry of heavy elements. Similarly, other explanations, including contamination by
a supernova explosion, or the dredge-up of s-process elements in an asymptotic giant 
branch star fail primarily because of the low light-element abundances (especially carbon) 
and the normal iron abundance. 

\subsection{Variability}

The \ledz\ light curve was discussed in \S\,\ref{s:light}; most common explanations for 
the amplitude and frequencies were deemed not viable. An explanation in terms of non-radial 
g-mode oscillations is favoured on the grounds that more than one frequency is
present, and the amplitude is variable; the latter could be explained in terms of beating
between unresolved modes. However, pulsation is not favoured on the grounds that the 
non-radial orders would have to be very high in order to account for the periods, which are very 
long for hot subdwarfs. In particular they lie well beyond the theoretical g-mode cutoff 
boundary as articulated for white dwarfs by \citet{hansen85} and which corresponds to 
$P_g L \leq 3600 $\,s in the current case, where $L^2=\ell(\ell+1)$.
The absence of a known driving mechanism is also a factor, but with radiative levitation
in operation at the surface, an undiagnosed opacity bump in the  interior would be 
a likely possibility. Indeed, the excitation of g-modes in Wolf-Rayet stars  has been
attributed to a similar phenomenon \citep{townsend06}. In order to place the 
\ledz\ light curve into a wider context, it is worth comparing it with other variable
and peculiar hot subdwarfs. 

\paragraph*{\crimson}. Pulsations in chemically-peculiar helium-rich subdwarfs were first discovered
by \citet{ahmad05}, who discovered light variations on timescales of $\approx 1800$\,s in the intermediate helium-rich 
hot subdwarf \crimson\  ($\teff = 32\,500$\,K, $\lgcs = 5.4$, $\log n_{\rm He}/n_{\rm H} = -0.58$, \citep{ahmad03}). 
Fine analysis  showed the surface of \crimson\ to have $\sim4$\,dex overabundances of zirconium, strontium, yttrium and germanium, 
which as so far been attributed to the effects of selective radiative levitation in the stellar photosphere \citep{naslim11}.
Whilst the  light variations have subsequently been confirmed and interpreted as non-radial g-mode pulsations 
\citep{jeffery11.ibvs,green11},  there is  debate over the precise effective temperature \citep{naslim11,green11},
and no completely satisfactory driving mechanism has been identified. An argument that the $\epsilon$-mechanism is active 
remains to be tested \citep{bertolami11}.  
-- need to refer to green+11 log g as well ... might be woirth looking at $Pg$ for this case also. -- 

\paragraph*{KIC\,1718290 = (SDSS)\,J192300+371504}  is a blue horizontal-branch star or `cool' hot subdwarf 
 ($\teff = 22\,100$\,K, $\lgcs = 4.72$)  with a super-solar helium abundance ($\log n_{\rm He}/n_{\rm H} = –0.45$) \citep{ostensen12}. 
It was observed with {\it Kepler} as an exoplanet target,  revealing a rich spectrum of low-amplitude pulsation modes with 
periods between one and twelve hours, most of which follow a regular spacing of 276.3 s. These fall into the classical 
range for g-mode non-radial pulsations \citep{jeffery06b}. Although having a 12\,h period and super-solar helium abundance
in common, the large difference in $\teff$ means that this star provides negligible insight for the present study. 

\paragraph*{KIC\,10449976 = (2MASS)\,J184714+474146}
is an extremely helium-rich subdwarf ($\teff = 40\,000$ K, $\lgcs = 5.3$ and $\log n_{\rm He}/n_{\rm H} = +1.2$) \citep{jeffery13a}.
{\it Kepler} photometry (quarters 3 and 5--9) shows evidence for a periodic modulation on a time-scale of $\approx3.9$\,d, but with variable amplitude.
Radial-velocity measurements over a 5\,d time-scale show an upper variability limit of $\approx50\pm20 \kms$. The origin of this
modulation remains to be confirmed; \citet{bear14} argue for reflection from a weather-affected planet but fail to show how such a
model can lose phase coherence over an interval of 160 or more orbits. Follow-up observations are yet to be carried out.  

\paragraph*{KIC\,9408967 = (SDSS)\,J19352+4555.}
\citet{ostensen10} identify three He-sdOB stars in the {\it Kepler} field: 
(Galex) J19034+3841, (SDSS) J19352+4555 and J19380+4649.
 None pulsate, but \citet{ostensen10} reported rapid drops in the {\it Kepler} Quarter 2 photometry of J19352+4555
with no apparent regularity. The maximum light variations in the other two He-sdOBs at low frequencies 
($100-500\,\mu$Hz) are 117 parts per million (J19034+3481) and 29 parts per million (J19380+4649).
 J19352+4555 is reported to be an extremely He-rich sdO star with (\teff,\logg) similar to
that of \ledz\ (\O{}stensen 2016, private communication).  
Reprocessed {\it Kepler}  light curves of J19352+4555  from Quarters 2 (short cadence and long cadence) and 
10 (long cadence only)  were extracted from the  {\it Kepler} archive. Examination of these shows no evidence for periodic 
short-term variability with an amplitude greater than 60 parts per million (2.5 $\sigma$). 
\paragraph*{V499\,Ser =  (SDSS)\, J160043.6+074802.9} was found by \citet{woudt06} to be a
very rapid hot sdO pulsator, with pulsation periods in the range 60 -- 120\,s. With
$\teff = 68\,500\pm1770$\,K, $\lgcs= 6.09\pm0.07$, and $\log n_{\rm He}/n_{\rm H} = -0.64\pm0.05$
\citep{latour11}, it is much hotter than \ledz, but the oscillation periods are compatible with being p-modes.
\citet{fontaine08b} argue that, as for the cooler sdB pulsators, radiative levitation can accumulate sufficient iron
and other high-opacity species to drive low-order low-degree p-mode pulsations in some hot subdwarfs 
with \teff\ between 60\,000 and 80\,000 K by the  $\kappa$-mechanism.  Again,  the surface abundance of 
helium is super-solar and similar to that in \ledz. 

\paragraph*{EO\,Cet = PB\,8783} was originally identified as a classical sdB pulsator \citep{koen97} having
 periods in the range 120 -- 135\,s and  an early F-type companion. 
Subsequent  spectroscopy showed \lineA{He}{ii}{4686} to be present, indicating
\teff\ well in excess of 36\,000\,K; this star may consequently have been the first pulsating sdO star 
discovered \citep{ostensen12b}. With similarities to V499\,Ser, further spectroscopy is required.  

\paragraph*{$\omega$ Cen} has been found to show a pulsation instability strip containing five hot subdwarfs
with multi-periodic oscillations with periods between 85 and 125\,s and with \teff\ between 48\,000 and 54\,000\,K
\citep{randall16}.  All are helium-poor. These p-mode pulsations are consistent with theoretical predictions and a
consequence of radiatively-driven diffusion in the interior enhancing the the iron-bump opacities, just as in the case
of V499\,Ser. Six short-period (85 -- 150\,s) pulsating hot subdwarfs have been detected in the globular cluster 
NGC\,2808 \citep{brownt13}, but are not strongly constrained in \teff. 

Thus there is evidence of both  p- and g-mode pulsations in helium-poor and helium-rich  hot subdwarfs
with $\teff > 30\,000$\,K. 
There is evidence of irregular and longer-term (3.9\,d) light variations in at least one 
case (KIC\,10449976), which has yet to be explained. 
One blue horizontal-branch star (KIC\,1718290) shows g-mode pulsation periods of up to 12\,h; these
can be reconciled with pulsation theory. 
There is no previous observational evidence or theoretical support for regular pulsations with 
periods as long as 11 -- 13 hours in any hot subdwarf. Hence, although pulsation remains the
preferred explanation for light variability in \ledz, it is likely to remain contentious without
additional evidence.  


\section{Conclusion}

We have obtained {\it NOT/ALFOSC} and {\it Subaru/HDS} spectroscopy, and {\it K2} photometry 
of the hot subdwarf \ledz, which we have augmented with archival {\it IUE} spectrophotometry and
broad-band photometry.  

{\it NOT/ALFOSC} spectroscopy rules out any short-period ($<7$\,d) binary companion, and 
provides fundamental atmospheric parameters of $\teff=38\,900\pm270$\,K, $\lgcs=5.97\pm0.11$ 
and $\log n_{\rm He}/n_{\rm H}=-0.57\pm0.01$ (formal errors). 
These results are supported by the {\it IUE} spectrophotometry which, in addition, give $E_{B-V}\approx0.03$
and angular radius $\theta = 1.062\pm0.006\times10^{-11}$ radians. 

{\it Subaru/HDS} spectroscopy provides elemental abundances in the stellar atmosphere 
for 14 species including yttrium,  germanium and lead, and upper limits for three, including 
carbon and iron. Carbon is at least 2 dex sub-solar and iron is approximately solar. Otherwise all measured
elements heavier than argon are $\approx 2 - 4$ dex overabundant,  whilst light elements
from nitrogen to sulphur are solar or sub-solar. \ledz\ thus forms the fourth member of a group
of ``heavy metal'' hot subdwarfs; these have intermediate helium-rich atmospheres and 3 to 4 dex 
overabundances of one or more of germanium, strontium, yttrium, zirconium, and lead. 
The peculiar chemistry probably arises as a result of radiative levitation producing a heavily stratified
atmosphere with overabundant species having a high specific opacity at the local temperature 
of the line-forming region.  Other interpretations are possible, but unlikely. Even when the chemistry
is understood, assessing the evolutionary origin of \ledz\ will present a significant challenge.  

Over 150 unidentified absorption lines in the {\it Subaru/HDS} spectrum present another challenge,
possibly leading to the identification of additional species. Other heavy-metal subdwarfs analyzed so 
far are essentially free of this problem. A high-resolution ultraviolet spectrum 
is well within reach of this UV-bright star using the {\it Hubble Space Telescope} and would provide  
crucial information about species not seen in the optical, about  stratification in the photosphere, 
and about the opacity required to model the photosphere correctly. 

{\it K2} photometry reveals a unique light curve having a dominant period around 10.8\,h, but with a
semi-amplitude which varies from 1 -- 5 ppt over the 75\,d of observation. The first harmonic of 
the 10.8\,h period, and another period at 13.3\,h are also present for most of the light curve. 
The periods are not constant throughout the observations, and may be affected by additional 
unresolved components. The favoured explanation for the light variability in \ledz\ is a multi-periodic
non-radial oscillation due to g-modes with very high radial order. However, the periods are 
some 300 times longer than the fundamental radial mode for such stars, and no theory 
predicts such modes to exist. Alternative explanations involving a stellar mass (or other) 
companion fail for lack of radial-velocity evidence. High-precision ($\sim 1\kms$) radial-velocity studies over
a 12 hour interval would be instructive.

Taking the extraordinary combination of observed properties and the associated difficulties in 
interpretation, \ledz\ presents one of the most challenging hot subdwarfs known to date and
 is a strong candidate to be  the first ``pulsating lead-rich hot subdwarf''.


\section*{Acknowledgments}
Research at the Armagh Observatory and Planetarium is supported by a grant-in-aid 
from the Northern Ireland Department for Communities.
CSJ acknowledges support from the UK Science and Technology Facilities Council (STFC) Grant No. 
ST/M000834/1.
HPP  acknowledges support from STFC Grant No. ST/M502268/1.
ASB gratefully acknowledges financial support from the Polish National Science Centre under project No.\,UMO-2011/03/D/ST9/01914.
This paper is based upon work supported by the National Auronautics and Space Adminstration under Grant NNH14ZDA001n-K2GO1.
Funding for this research was provided by the National Science Foudation (USA) grant \#1312869.

This paper includes data collected by the {\it Kepler} mission.  Funding for the {\it Kepler} mission is 
provided by the NASA Science Mission directorate.

This paper is based in part on data collected at Subaru Telescope, which is operated by the National Astronomical Observatory of Japan.

Based on observations made with the Nordic Optical Telescope, operated by the Nordic Optical Telescope Scientific Association at the
 Observatorio del Roque de los Muchachos, La Palma, Spain, of the Instituto de Astrofisica de Canarias.

Some of the data presented in this paper were obtained from the Mikulski Archive for Space Telescopes
(MAST). STScI is operated by the Association of Universities for Research in Astronomy, Inc., under NASA 
contract NAS5-26555. Support for MAST for non-HST data is provided by the NASA Office of Space
Science via grant NNX09AF08G and by other grants and contracts.

This research has made use of the SIMBAD database, operated at CDS, Strasbourg, France.

This work has made use of the Vienna Atomic Line Database (VALD) database, operated at Uppsala University, 
the Institute of Astronomy of the Russian Academy of Sciences in Moscow, and the University of Vienna, 
the Atomic Line List, hosted by the Department of Physics and Astronomy, University of Kentucky,
and the National Institute of Standards and Technology (NIST) Atomic Spectra Database, which is hosted by the U.S. Dept of Commerce. 

The TheoSSA {\sc tmap} service (http://dc.g-vo.org/theossa) used to retrieve theoretical spectra for this paper was
 constructed as part of the activities of the German Astrophysical Virtual Observatory. 

This research has made use of {\sc iraf},  the Image Reduction and Analysis Facility,  
written and supported by the National Optical Astronomy Observatories (NOAO) in Tucson, Arizona. 
NOAO is operated by the Association of Universities for Research in Astronomy (AURA), Inc. under 
cooperative agreement with the National Science Foundation.

This research has made use of the period-analysis software {\sc ts} \citep{templeton04} 
 made available by the AAVSO, Cambridge, Massachusetts, USA. 

The authors are grateful to Thomas Rauch for assistance with running {\sc tmap},  to
Philip Hall for providing the zero-age helium main-sequence data for Fig.\,\ref{f:tg}, 
and to the referee for drawing attention to numerous details which deserved correction
or improvement.


\bibliographystyle{mnras}
\bibliography{ehe}

\appendix
\renewcommand\thefigure{A.\arabic{figure}} 
\renewcommand\thetable{A.\arabic{table}} 

\section[]{Spectral Atlas and line abundances for \ledz}
\label{s:app1}
\label{s:lines}
\begin{figure*}
\epsfig{file=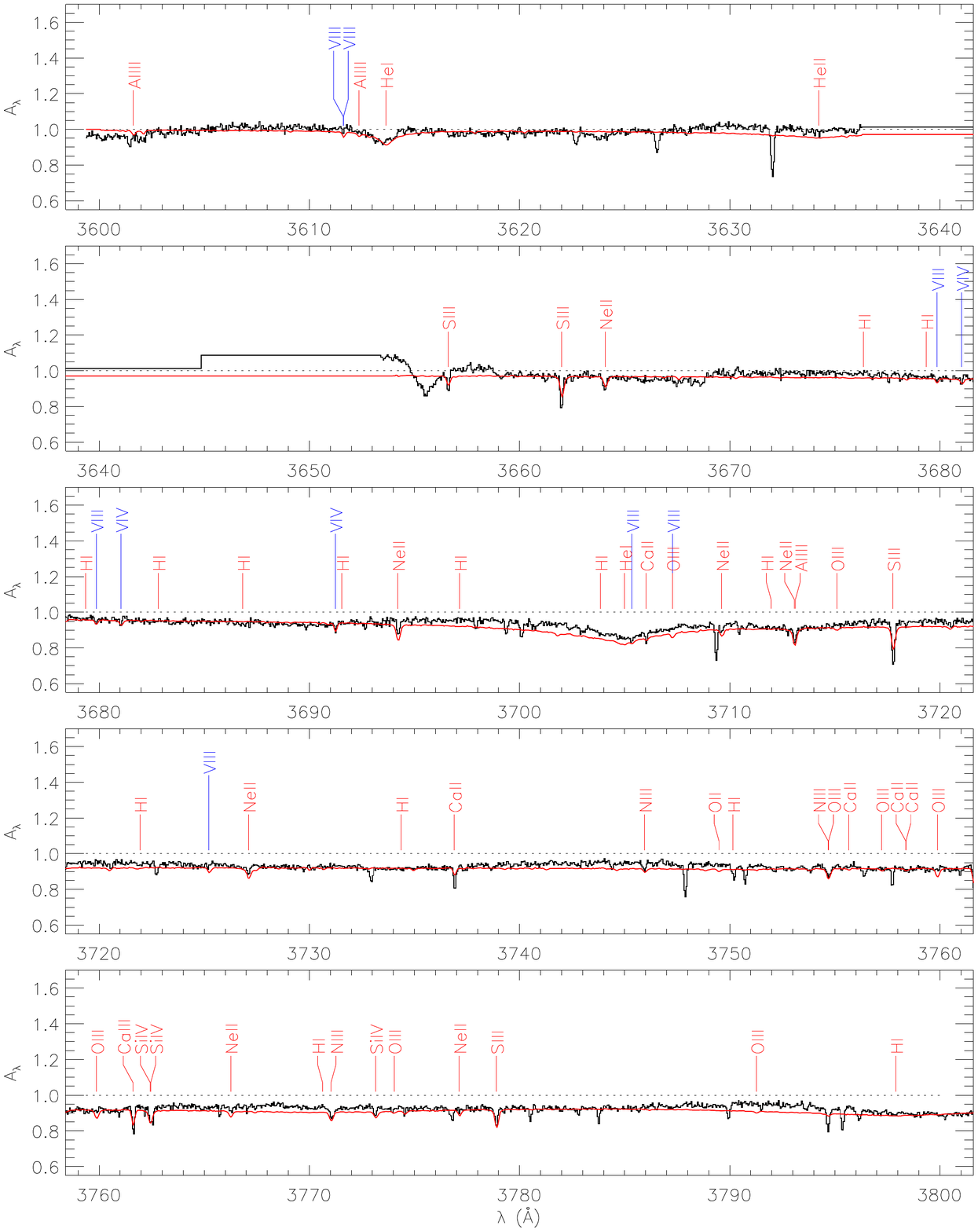,width=1.0\textwidth}
\caption{{\it Subaru/HDS} spectrum of \ledz\ (black histogram), and best-fit model
having  $\teff=39\,000$\,K, $\lgcs=6.0$),  $n_{\rm He}/n_{\rm H}=0.25$ 
and abundances shown in Table~\ref{t:abunds}.
Lines with theoretical equivalent widths greater than $5$m\AA\ are identified wherever possible. Gaps in 
the observed spectrum correspond to major instrumental artefacts. 
}
\label{f:atlasA}
\end{figure*}

\begin{figure*}
\epsfig{file=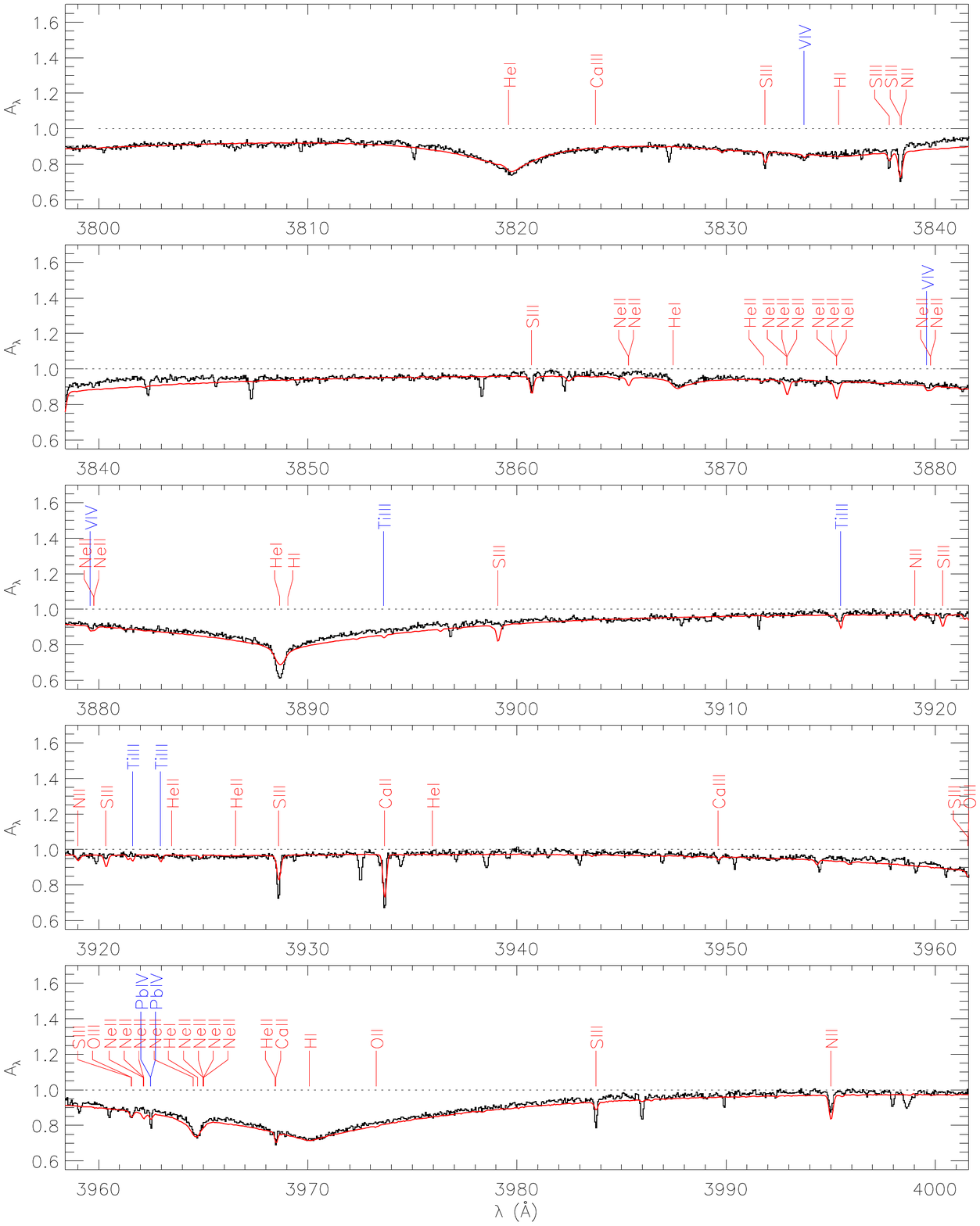,width=1.0\textwidth}
\caption{As Fig.~\ref{f:atlasA} (contd.)}
\label{f:atlasB}
\end{figure*}

\begin{figure*}
\epsfig{file=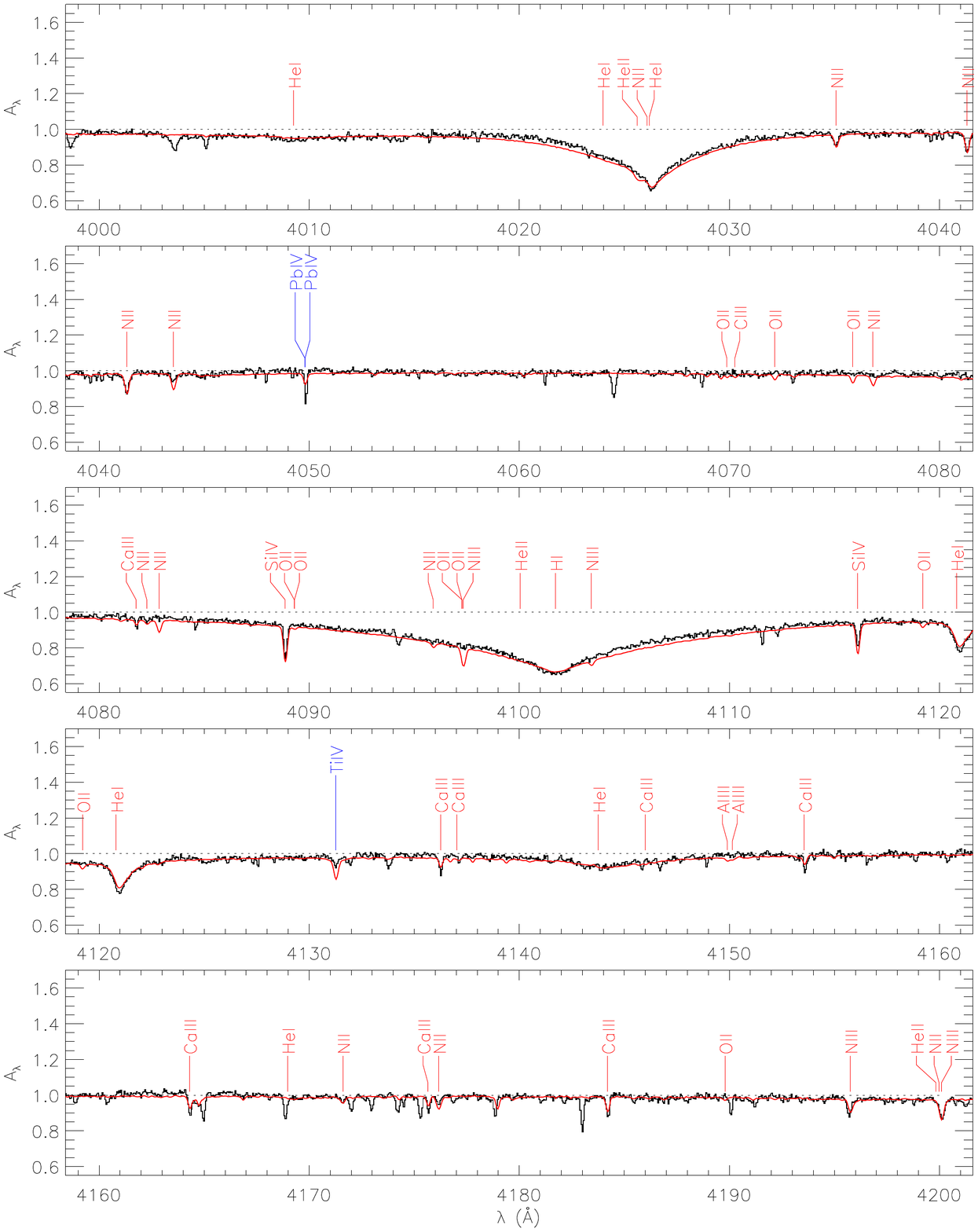,width=1.0\textwidth}
\caption{As Fig.~\ref{f:atlasA} (contd.)}
\label{f:atlasC}
\end{figure*}

\begin{figure*}
\epsfig{file=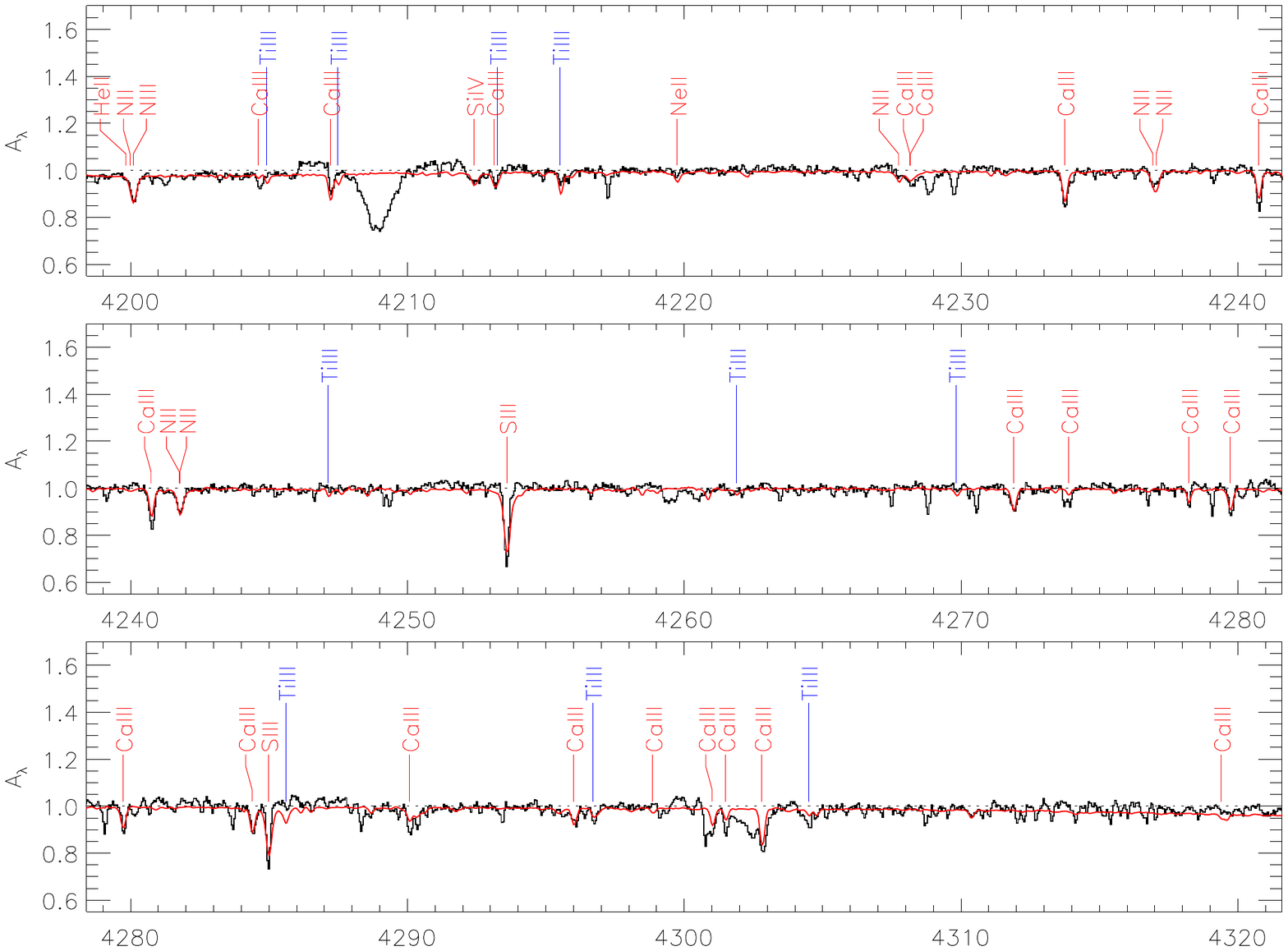,width=1.0\textwidth}
\caption{As Fig.~\ref{f:atlasA} (contd.)}
\label{f:atlasD}
\end{figure*}

\begin{figure*}
\epsfig{file=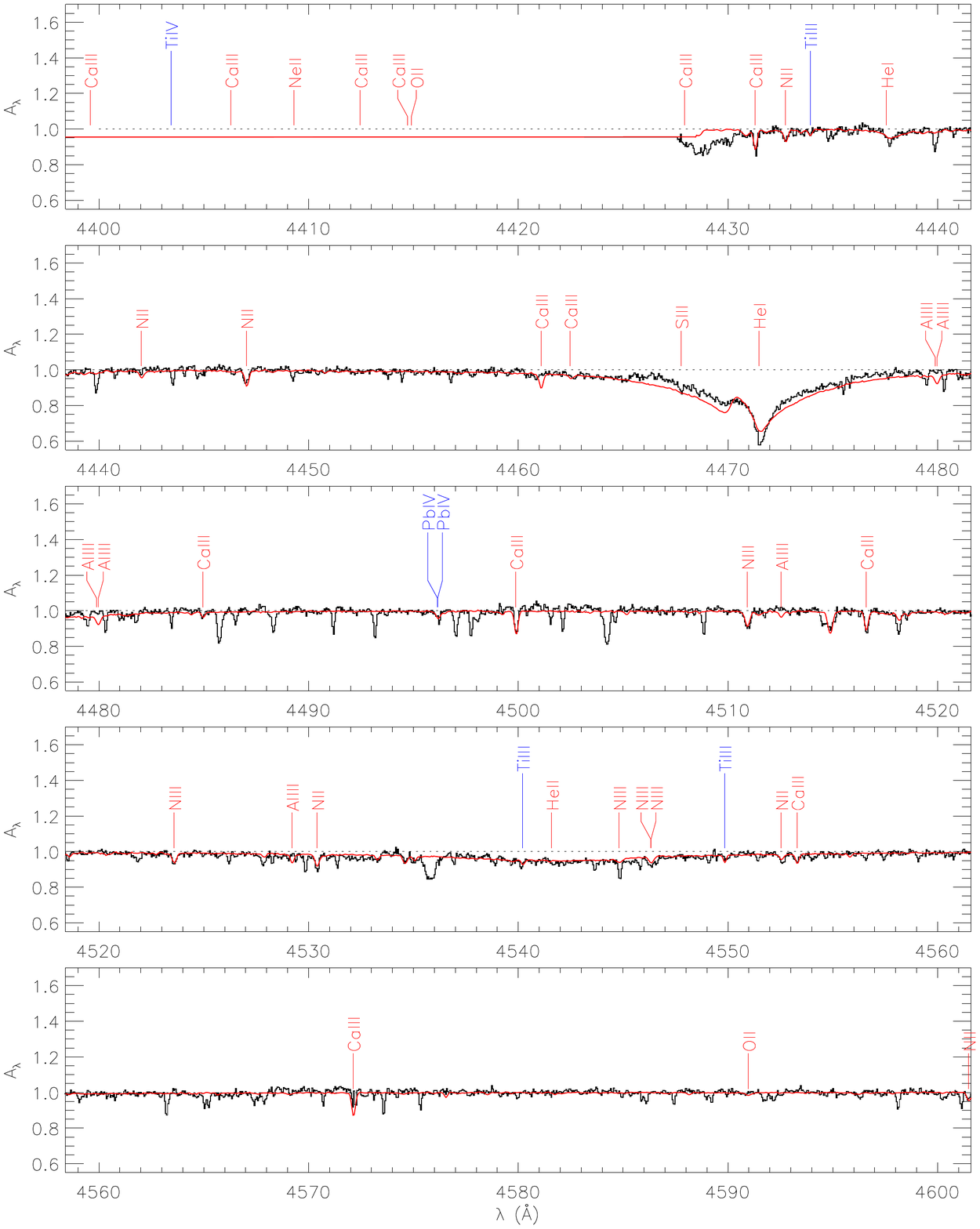,width=1.0\textwidth}
\caption{As Fig.~\ref{f:atlasA} (contd.)}
\label{f:atlasE}
\end{figure*}

\begin{figure*}
\epsfig{file=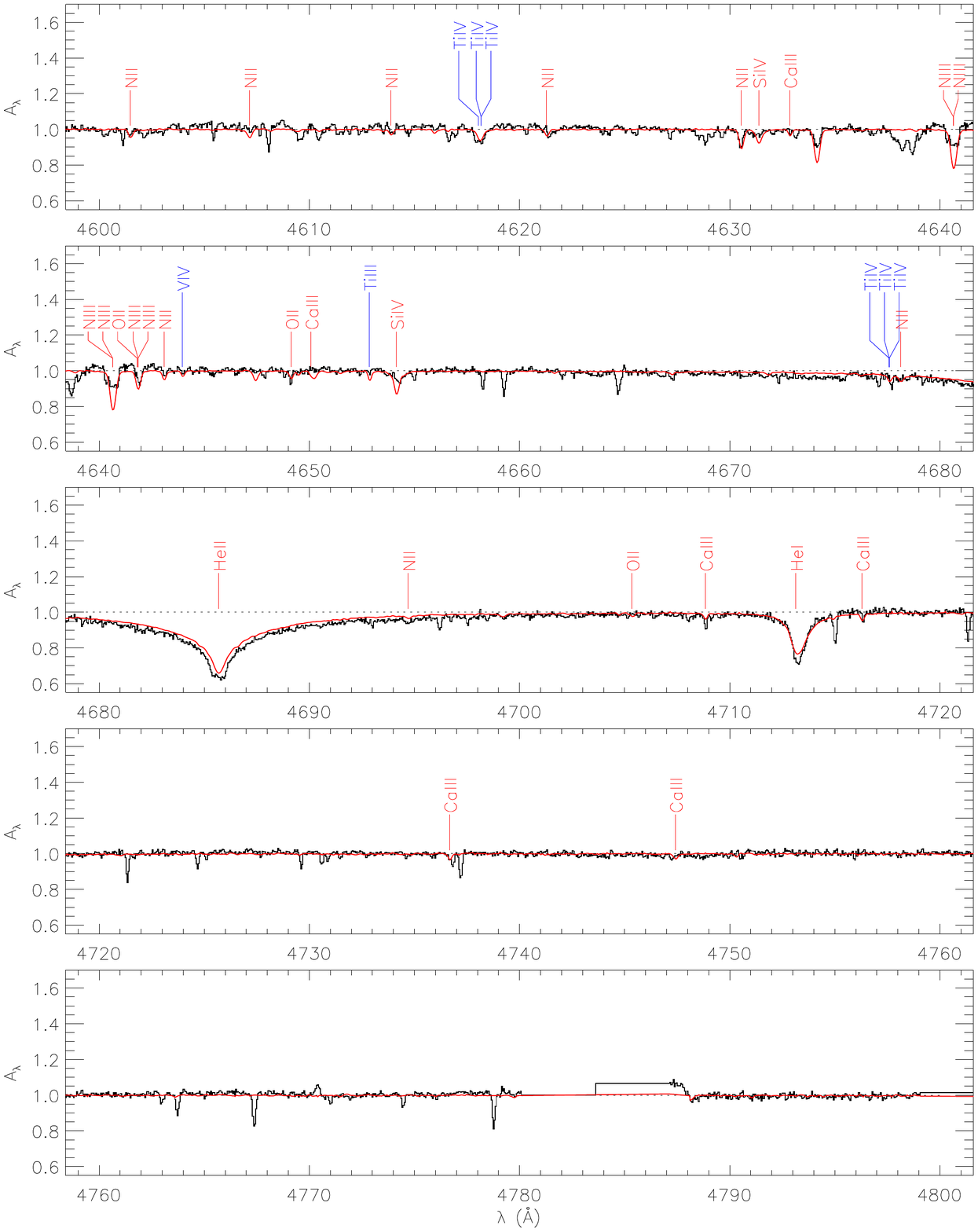,width=1.0\textwidth}
\caption{As Fig.~\ref{f:atlasA} (contd.)}
\label{f:atlasF}
\end{figure*}

\begin{figure*}
\epsfig{file=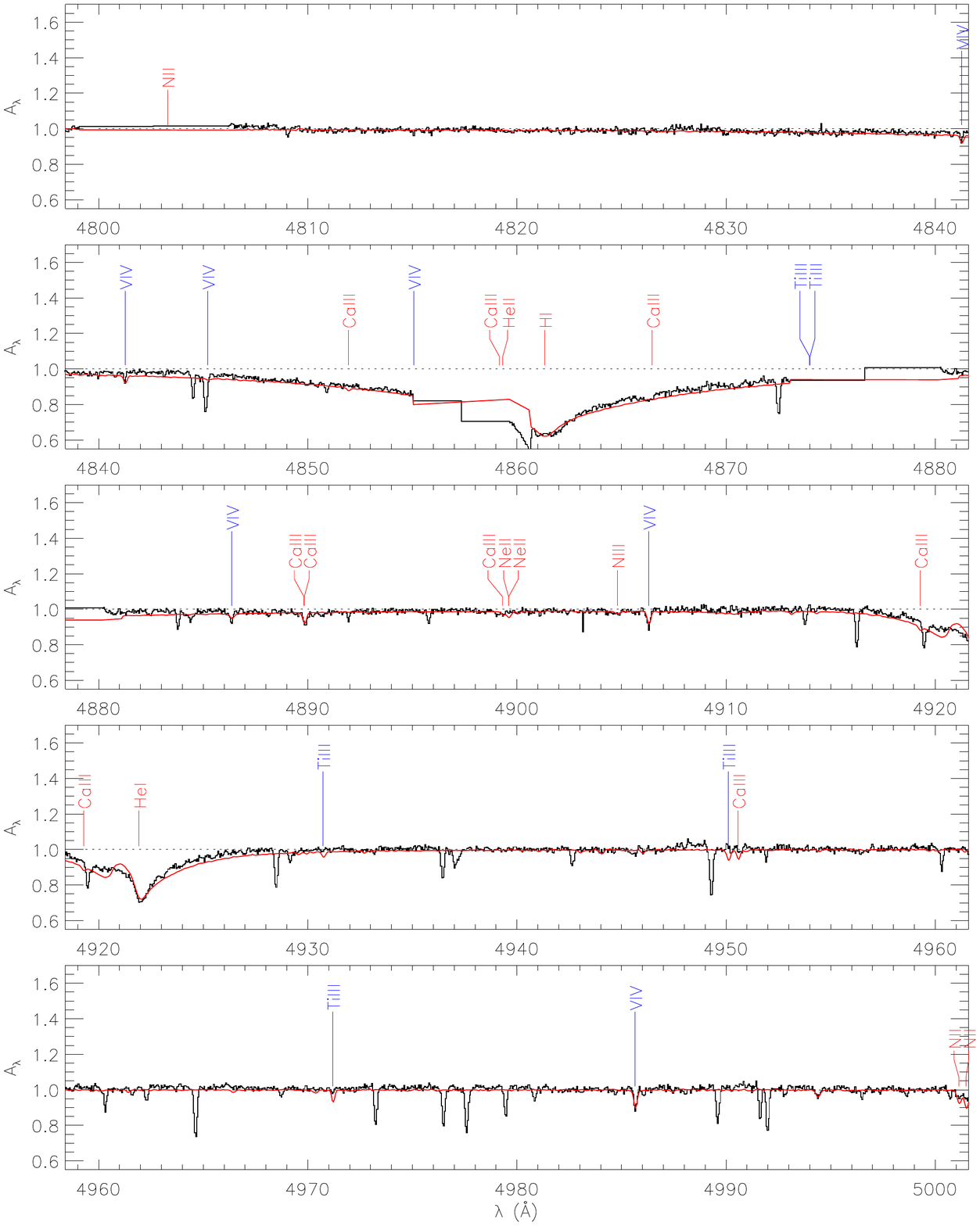,width=1.0\textwidth}
\caption{As Fig.~\ref{f:atlasA} (contd.)}
\label{f:atlasG}
\end{figure*}

\begin{figure*}
\epsfig{file=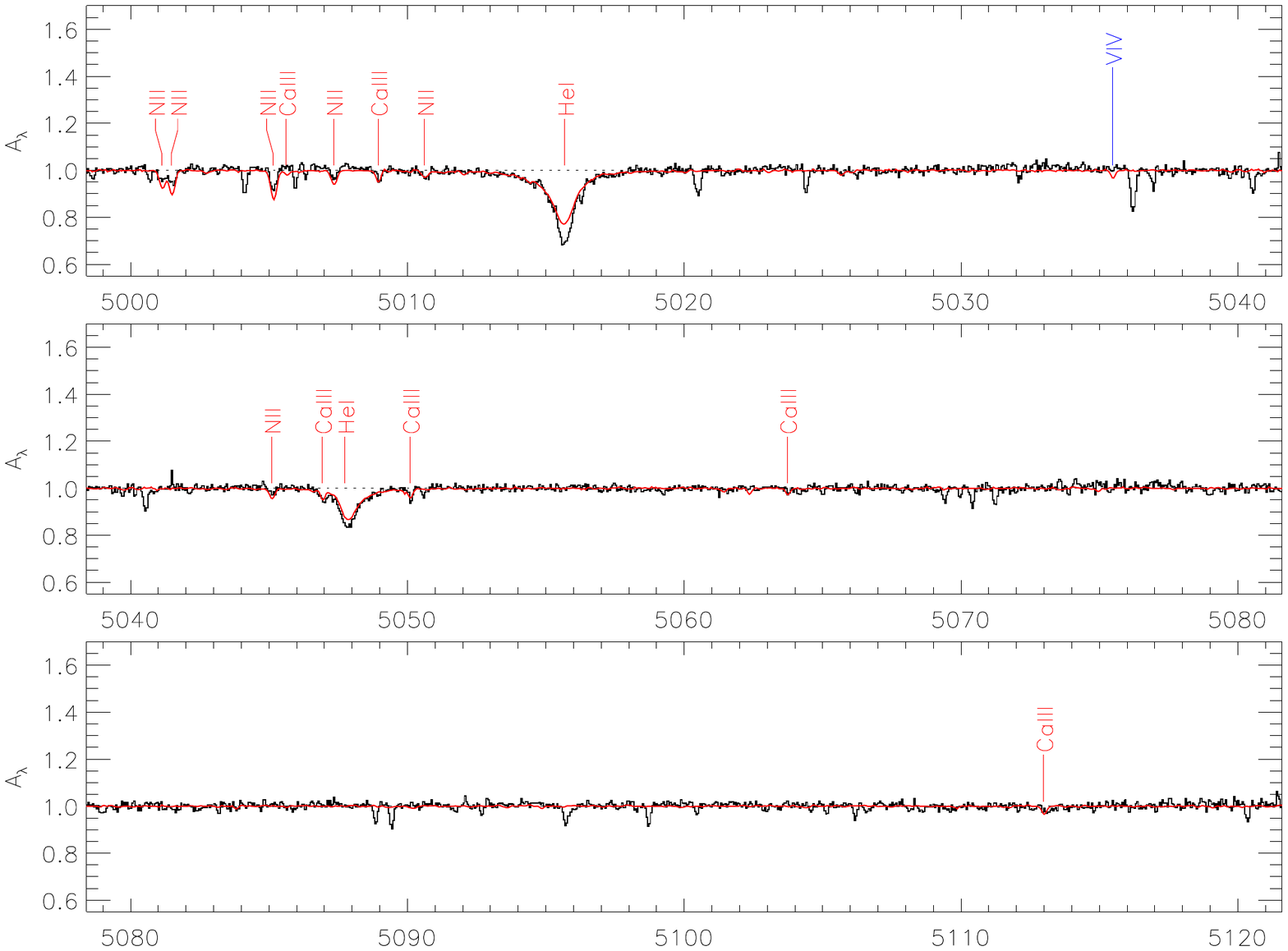,width=1.0\textwidth}
\caption{As Fig.~\ref{f:atlasA} (contd.)}
\label{f:atlasH}
\end{figure*}

Figures \ref{f:atlasA} to \ref{f:atlasH} contain an atlas of the {\it Subaru/HDS} spectrum of \ledz,
with the best model fit and identifications of absorption lines. Table \ref{t:lines} shows equivalent 
widths, adopted transition probabilities, and derived abundances for individual lines in the
above spectrum.  

\begin{table}
\caption{Equivalent widths and scaled abundances $\log \epsilon$ for 
identified unblended absorption lines. 
The sources of oscillator strengths ($gf$) are given with the ion designation.
Abundances were computed assuming a model atmosphere with $\teff=39\,000$\,K,
$\lgcs=6.0$, $\vt=2\kms$, and composition as in Table\,\ref{t:abunds}.
Errors on the mean abundance per ion are given as the standard
deviation where the number of lines $n>2$, the semi-range where $n=2$
and adopting the mean measurement error of $\pm2$\,m\AA\ where $n=1$. 
} 
\label{t:lines}
\begin{flushleft}
\begin{tabular}{lrrll}
\hline
\noalign{\smallskip}
Ion &  Reference \\
$\lambda/$\AA\ &$\log{gf}$ & \ew\,/{\rm m\AA} & $\log \epsilon$\\
\hline
\ion{N}{ii} &  \multicolumn{3}{l}{\citet{becker89}} \\
 3919.01 & -0.335 &   8 & 8.13 & \\
 3995.00 &  0.225 &  25 & 7.88 & \\
 4171.60 &  0.280 &   7 & 7.87 & \\
 4175.66 & -1.180 &  11 & 7.79 & \\
 4241.18 & -0.336 &  29 & 8.01 & \\
 4431.82 & -0.152 &  12 & 7.80 & \\
 4432.74 &  0.595 &   7 & 7.54 & \\
 4447.03 &  0.238 &  23 & 8.14 & \\
 4530.40 &  0.671 &  19 & 8.07 & \\
 4552.53 &  0.207 &  19 & 8.52 & \\
 4601.48 & -0.385 &  13 & 8.18 & \\
 4607.16 & -0.483 &   7 & 8.00 & \\
 4613.87 & -0.607 &  14 & 8.45 & \\
 4630.54 &  0.093 &  23 & 8.00 & \\
 4643.09 & -0.385 &  13 & 8.18 & \\
 5001.13 &  0.282 &  17 & 7.60 & \\
 5005.15 &  0.612 &  24 & 7.89 & \\
 5007.33 &  0.161 &  18 & 8.21 & \\
 5045.09 & -0.389 &   7 & 7.89 & \\
 \cmidrule{4-5}
         &  & & \multicolumn{2}{l}{$8.01\pm0.25$} \\
 \cmidrule{4-5}
\ion{N}{iii} &  \multicolumn{3}{l}{\citet{butler84}} \\
 3745.95 & -0.780 &   8 & 8.03 & \\
 3754.69 & -0.480 &  17 & 8.14 & \\
 3771.03 & -0.300 &  16 & 7.93 & \\
 4544.80 & -0.143 &  17 & 8.33 & \\
 4546.32 &  0.017 &  18 & 8.32 & \\
 4641.85 & -0.815 &  29 & 8.20 & \\
 \cmidrule{4-5}
         &  & & \multicolumn{2}{l}{$8.16\pm0.16$} \\
 \cmidrule{4-5}
\ion{O}{ii} &  \multicolumn{3}{l}{\citet{bell94}} \\
 4649.14 &  0.308 &   9 & 7.47 & \\
 \cmidrule{4-5}
         &  & & \multicolumn{2}{l}{$7.47\pm0.09$} \\
 \cmidrule{4-5}
\ion{O}{iii} & \\
 3754.70 & -0.099 &  12 & 7.38 & \\
 \cmidrule{4-5}
         &  & & \multicolumn{2}{l}{$7.38\pm0.07$} \\
 \cmidrule{4-5}
\ion{Ne}{ii} &  \multicolumn{3}{l}{\citet{wiese66}} \\
 3664.07 & -0.260 &  16 & 7.75 & \\
 3709.62 & -0.330 &   7 & 7.44 & \\
 3713.08 &  0.260 &  14 & 7.26 & \\
 \cmidrule{4-5}
         &  & & \multicolumn{2}{l}{$7.48\pm0.25$} \\
 \cmidrule{4-5}
\ion{Mg}{ii} & \multicolumn{3}{l}{\citet{wiese69}} \\
 4481.13 &  0.568 &   7 & 6.25 & \\
 \cmidrule{4-5}
         &  & & \multicolumn{2}{l}{$6.25\pm0.11$} \\
 \cmidrule{4-5}
\ion{Al}{iii} & \multicolumn{3}{l}{\citet{cunto93}} \\
 3601.63 &  0.000 &  12 & 6.47 & \\
 \cmidrule{4-5}
         &  & & \multicolumn{2}{l}{$6.47\pm0.07$} \\
 \cmidrule{4-5}
\hline
\end{tabular}
\end{flushleft}
\end{table}

\begin{table}
\addtocounter{table}{-1}
\caption{(cont.)}
\begin{flushleft}
\begin{tabular}{lrrll}
\hline
\noalign{\smallskip}
Ion &  Reference \\
$\lambda/$\AA\ &$\log{gf}$ & \ew\,/{\rm m\AA} & $\log \epsilon$\\
\hline
\ion{Si}{iv} &  \multicolumn{3}{l}{\citet{becker90}} \\
 3762.43 &  0.250 &  22 & 6.57 & \\
 3773.15 & -0.010 &   6 & 6.28 & \\
 4088.85 &   0.199 &   37 & 6.14  \\
 4116.10 & -0.103 &  32 & 6.33 & \\
 4631.38 &  1.217 &   6 & 5.94 & \\
 4654.14 &  1.486 &  24 & 6.31 & \\
 \cmidrule{4-5}
         &  & & \multicolumn{2}{l}{$6.26\pm0.21$} \\
 \cmidrule{4-5}
\ion{S}{iii} &  \multicolumn{3}{l}{\citet{hardorp70}} \\
 3656.60 & -0.830 &  13 & 7.66 & \\
 3662.01 & -0.380 &  25 & 7.56 & \\
 3717.77 & -0.190 &  42 & 7.65 & \\
 3778.90 & -0.290 &  18 & 7.28 & \\
 3837.80 & -0.570 &  21 & 7.65 & \\
 3928.59 & -0.190 &  43 & 7.70 & \\
 3961.56 & -0.810 &  11 & 7.58 & \\
 3983.77 & -0.720 &  26 & 7.94 & \\
 4253.59 &  0.400 &  70 & 7.42 & \\
 4284.98 &  0.110 &  50 & 7.48 & \\
 4332.69 & -0.240 &  50 & 7.83 & \\
 \cmidrule{4-5}
         &  & & \multicolumn{2}{l}{$7.61\pm0.18$} \\
 \cmidrule{4-5}
\ion{Cl}{iii} &  \multicolumn{3}{l}{\citet{wiese69}} \\
 3720.45 &  0.350 &   7 & 6.34 & \\
 \cmidrule{4-5}
         &  & & \multicolumn{2}{l}{$6.34\pm 0.11$} \\
 \cmidrule{4-5}
\ion{Ti}{iii} &  \multicolumn{3}{l}{\citet{warner69}} \\
 3915.26 &  0.066 &  14 & 8.01 & \\
 4214.93 & -0.189 &   9 & 8.06 & \\
 \cmidrule{4-5}
         &  & & \multicolumn{2}{l}{$8.03\pm0.03$} \\
 \cmidrule{4-5}
\ion{Ti}{iv} &  \multicolumn{3}{l}{\citet{kurucz99}} \\
 4131.26 &  0.918 &  14 & 6.68 & \\
 4618.04 &  0.277 &  23 & 7.34 & \\
 4677.59 &  0.339 &  12 & 7.38 & \\
 \cmidrule{4-5}
         &  & & \multicolumn{2}{l}{$7.13\pm0.39$} \\
 \cmidrule{4-5}
\ion{V}{iv} &  \multicolumn{3}{l}{\citet{martin88}} \\
 4841.26 &  0.150 &   7 & 7.16 & \\
 4906.29 &  0.300 &  19 & 7.57 & \\
 4985.64 &  0.520 &  21 & 7.45 & \\
 5130.78 &  0.620 &  38 & 7.85 & \\
 5146.52 &  0.410 &  20 & 7.50 & \\
 \cmidrule{4-5}
         &  & & \multicolumn{2}{l}{$7.51\pm0.25$} \\
 \cmidrule{4-5}
\ion{Ge}{iii} & \multicolumn{3}{l}{\citet{naslim11}} \\
 4178.96 &  0.341 &  19 & 6.24 & \\
 \cmidrule{4-5}
         &  & & \multicolumn{2}{l}{$6.24\pm0.04$} \\
 \cmidrule{4-5}
\ion{Y}{iii} & \multicolumn{3}{l}{\citet{naslim11}} \\
4039.600 & 1.005 & 8   & 5.47 & \\
4040.110 & 1.005 & 6   & 5.28 & \\
 \cmidrule{4-5}
         &  & & \multicolumn{2}{l}{$5.37\pm 0.09$} \\
 \cmidrule{4-5}
\ion{Pb}{iv} & \multicolumn{3}{l}{\citet{naslim13}} \\
 3962.48 & -0.025 &  12 & 5.30 & \\
 4049.80 & -0.010 &  24 & 5.66 & \\
 4496.15 & -0.237 &  10 & 5.51 & \\
 \cmidrule{4-5}
         &  & & \multicolumn{2}{l}{$5.49\pm0.18$} \\
 \cmidrule{4-5}
\hline
\end{tabular}
\end{flushleft}
\end{table}

\begin{table}
\addtocounter{table}{-1}
\caption{(cont.)}
\begin{flushleft}
\begin{tabular}{lrrll}
\hline
\noalign{\smallskip}
Ion &  Reference \\
$\lambda/$\AA\ &$\log{gf}$ & \ew\,/{\rm m\AA} & $\log \epsilon$\\
\hline

\ion{Ca}{ii} &  \multicolumn{3}{l}{\citet{wiese69}} \\
 3706.02 & -0.441 &   6 & 8.15 & \\
 3933.66 &  0.134 &  51 & 7.95 & \\
 \cmidrule{4-5}
         &  & & \multicolumn{2}{l}{$8.05\pm0.10$} \\
 \cmidrule{4-5}
\ion{Ca}{iii} &  \multicolumn{3}{l}{\citet{kurucz99}} \\
 3706.02 & -0.441 &   6 & 8.15 & \\
 3761.61 & -1.364 &  18 & 7.95 & \\
 3949.61 & -0.748 &   8 & 8.40& \\
 4136.25 &  0.166 &  22 & 8.47 & \\
 4136.25 &  0.166 &   9 & 7.94 & \\
 4153.57 & -0.334 &  22 & 8.54 & \\
 4164.30 &  0.258 &  26 & 8.36 & \\
 4175.65 & -0.359 &  17 & 8.44 & \\
 4184.20 & -0.045 &  21 & 8.23 & \\
 4211.61 & -0.506 &   6 & 8.41 & \\
 4213.13 & -0.394 &  13 & 8.27 & \\
 4233.71 & -1.092 &  45 & 8.50 & \\
 4256.65 & -0.473 &   7 & 8.43 & \\
 4271.82 & -0.919 &  25 & 8.25 & \\
 4273.88 & -0.654 &  14 & 8.63 & \\
 4278.22 & -0.369 &  12 & 8.26 & \\
 4278.82 & -1.028 &  29 & 8.32 & \\
 4283.56 & -2.455 &  33 & 8.31 & \\
 4301.01 &  0.401 &  33 & 8.62 & \\
 4431.29 &  0.273 &  19 & 7.92 & \\
 4462.47 &  0.120 &   7 & 8.11 & \\
 4484.40 & -0.828 &   7 & 8.54 & \\
 4499.88 &  0.472 &  31 & 8.18 & \\
 4516.59 &  0.287 &  27 & 8.24 & \\
 4553.29 &  0.049 &  16 & 8.30 & \\
 4716.29 & -0.439 &   9 & 8.26 & \\
 4736.67 & -0.116 &   9 & 8.26 & \\
 4889.82 & -0.232 &  17 & 8.75 & \\
 4899.31 & -0.322 &   5 & 8.17 & \\
 4919.28 &  0.057 &  30 & 8.79 & \\
 5046.92 & -0.002 &  15 & 8.44 & \\
 5050.09 & -0.290 &  10 & 8.26 & \\
 5112.98 & -0.123 &   8 & 8.20& \\
 5137.73 &  0.086 &  11 & 8.20 & \\
 \cmidrule{4-5}
         &  & & \multicolumn{2}{l}{$8.32\pm0.21$} \\
 \cmidrule{4-5}
\hline
\end{tabular}
\end{flushleft}
\end{table}

\label{lastpage}

\newpage

\end{document}